\documentclass{article}

\usepackage{arxiv}

\usepackage[utf8]{inputenc} 
\usepackage[T1]{fontenc}    
\usepackage{hyperref}       
\usepackage{url}            
\usepackage{booktabs}       
\usepackage{amsfonts}       
\usepackage{nicefrac}       
\usepackage{microtype}      
\usepackage{lipsum}		
\usepackage{graphicx}
\usepackage{natbib}
\usepackage{doi}

\title{Four-function generalization and separable structures of the Pleba{\'n}ski spacetime with sources}

\author{ \href{https://orcid.org/0009-0006-9671-3719}{\includegraphics[scale=0.06]{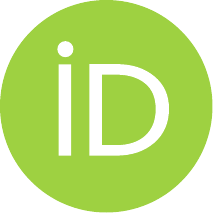}\hspace{1mm}Alfonso S.~Acevedo}\\
	Physics Department\\
	Cinvestav\\
	Mexico City, 2508\\
	\texttt{alfonso.acevedo@cinvestav.mx} \\
	\And
	\href{https://orcid.org/0000-0002-1237-7134}{\includegraphics[scale=0.06]{orcid.pdf}\hspace{1mm}Nora ~Breton} \\
	Physics Department\\
	Cinvestav\\
	Mexico City, 2508 \\
	\texttt{nora.breton@cinvestav.mx} \\
}

\renewcommand{\shorttitle}{}
\usepackage{graphicx}%
\usepackage{multirow}%
\usepackage{amsmath,amssymb,amsfonts}%
\usepackage{gensymb, upgreek, siunitx}
\usepackage{amsthm}%
\usepackage{mathrsfs}%
\usepackage[title]{appendix}%
\usepackage{xcolor}%
\usepackage{textcomp}%
\usepackage{manyfoot}%
\usepackage{booktabs}%
\usepackage{algorithm}%
\usepackage{algorithmicx}%
\usepackage{algpseudocode}%
\usepackage{listings}%
\usepackage{lmodern}
\usepackage{kpfonts}
\hypersetup{
pdftitle={Four-function generalization and separable structures of the Pleba{\'n}ski spacetime with sources},
pdfsubject={gr-qc},
pdfauthor={Alfonso S. Acevedo, N. Breton},
pdfkeywords={Integrability, stationary axisymmetric metrics, Pleba{\'n}sk metric, black holes},
}
\begin{document}
\maketitle

\begin{abstract}
We determine a four-function generalization of the Pleba{\'n}ski spacetime, depending on three arbitrary functions of the radial coordinate, and one function on the angular coordinate.  For the generalized Pleba{\'n}ski spacetime, we analyze the separability of the Hamilton-Jacobi equations, and the trajectories of a charged test particle are derived from the motion constants. 
The Klein-Gordon equation separability is established and the  Killing horizons are presented as well. Then we introduce  a conformal factor to the Pleba{\'n}ski metric and discuss the conditions that preserve the separability.  Finally we show a possible stress–energy tensor that may be the source of some of the generalized metrics.
\end{abstract}




\maketitle
\section{Introduction}\label{sec1}
Even before they were accepted as legitimate astrophysical objects \cite{BH}, black holes (BHs)  were used to explain astrophysical observations \cite{BHA}. The black hole (BH) creates a clear distinction between two regions of spacetime: the inner, inaccessible region and the outer region, that might be idealized as a vacuum spacetime but distorted by the presence of gravitation. One way in which information can be extracted from a BH is through the geodesics of particles and waves propagating in its neighbourhood.

The most important astrophysical solution to the Einstein equations is the Kerr solution, which describes a stationary, axially symmetric spacetime and hence can be a first approximation to a vacuum, exterior spacetime of a rotating, axially symmetric black hole or ultracompact object. The Kerr metric belongs to the family of Type-D metrics, distinguished by two obvious symmetries: stationarity and axisymmetry, associated with the two Killing vectors $\partial_{\tau}$ and $\partial_{\phi}$. 
\newline
The physically interesting solutions of Einstein-Maxwell equations of type D, that generalize Kerr-NUT with an electromagnetic field, with six continuous and one discrete parameter, were presented by J. Pleba{\'n}ski in \cite{Pleb}. Although a very wide class of solutions, it
was generalized by adding a conformal factor $\Omega = (1 - pq)^2$, by J. Pleba{\'n}ski and M. Demia{\'n}ski, constructing a family of solutions of type D of Einstein-Maxwell 
equations which contains seven continuous parameters, constituting the most general  family of Type D spacetimes with electromagnetic field and a non-zero cosmological constant \cite{PlebDem}. The general Pleba{\'n}ski-Demia{\'n}ski (PD) metric also admits the conformal generalization of the Killing-Yano tensor \cite{Kubiznak}, that in certain cases can be deduced by inspection of the form of the line element. We will focus in the non conformal case, ie, the Pleba{\'n}ski spacetime from \cite{Pleb}.

Additionally, these metrics possess hidden symmetries related to the existence of a Killing tensor, see \cite{FKK} for a review. The Killing tensor provides an additional constant of motion that in the Kerr metric is known as the Carter constant \cite{Car68}; the existence of a fourth motion constant allows us to integrate the geodesic equations.  These symmetries are also associated with the separability of the Hamilton-Jacobi and Klein-Gordon equations \cite{Btt}. The practical relevance of solving the Hamilton-Jacobi and Klein-Gordon equations lies in the description of the movement of particles and waves in the vicinity of black holes.
\newline
Recently in \cite{BaVis} a generalization of the Kerr metric was presented; inspired by that work we generalize in a similar way the Pleba{\'n}ski stationary axisymmetric metric solution of the Einstein-Maxwell equations \cite{Pleb}. The generalization of the metric functions leads to spacetimes that can be singularity-free; however that are no longer Einstein-Maxwell solutions neither algebraically degenerated (i.e. are no type D but algebraically general). In this paper we additionally present possible sources to the generalized metrics, which then makes them solutions of the Einstein equations coupled with matter sources that might be scalar fields and/or nonlinear electromagnetic fields. In this sense we should distinguish between models (off-shell), with no specified matter, and solutions of the Einstein equations with the appropriate matter (on-shell).

This article is organized as follows:
\newline
First, in Section 2, we summarize the general properties of the Pleba\'nski  spacetime and its symmetries to examine a generalized form of the Pleba{\'n}ski spacetime that preserves stationarity and axisymmetry. Then we introduce four generalizing functions: three depend solely on the radial coordinate, while the fourth depends on the angular coordinate $\theta$. The Pleba\'nski metric is recovered as a limiting case of these functions. The separability of the Hamilton–Jacobi (HJ) and Klein–Gordon (KG) equations is analyzed in Sections 2.1 and 2.2, respectively, and the equations of motion for a charged test particle are integrated using the conserved quantities.
In Section 3 the Killing horizons and the surface gravity of the generalized Pleba\'nski  metric are derived.
In Section 4, a conformal factor is introduced into the Pleba\'nski  metric, showing that, in general, the separability is not preserved. However, by imposing certain conditions on the analytic form of the conformal factor, it is possible to achieve separability of either the KG or the HJ equation, but not both with the same conformal factor.
\newline
Finally, in Section 5, a proposal is presented for the stress-energy tensor that can act as a source for static cases of the generalized Pleba\'nski  metrics. The stress-energy tensor considered is a scalar field and a nonlinear electromagnetic field, and some examples illustrate the wide spectra of solutions that the generalized functions can provide. The dominant energy condition is also addressed in the presented examples.

\section{Pleba{\'n}ski spacetime}\label{sect2}
The Pleba{\'n}ski metric in the Boyer coordinates $(\tau, \phi, p, q)$ is given by \cite{PlebDem}
\begin{equation}
    \textbf{g}=\left[ -\frac{Q(\textbf{d}\tau-\omega p^2\textbf{d}\phi)^2}{\Sigma} + \frac{P(\omega\textbf{d}\tau+q^2\textbf{d}\phi)^2}{\Sigma} + \frac{\Sigma}{P}\textbf{d}p^2 + \frac{\Sigma}{Q}\textbf{d}q^2\right].
\label{PD}
\end{equation}
where $P=P(p)$ and $Q=Q(q)$ are fourth degree polynomials,
\begin{equation*}
\begin{split}
Q(q) & =k+e^2+g^2-2mq+\epsilon q^2 - 2nq^3 - (k+\Lambda/3)q^4,\\
P(p) & =k+2np-\epsilon p^2 - 2mp^3 - (k+e^2+g^2+\Lambda/3)p^4,
\end{split}
\end{equation*}
and $\Sigma = q^2 + \omega^2 p^2$.
The parameters that characterize this solution are the electric charge $e$, magnetic charge $g$, cosmological constant $\Lambda$, rotation parameter $\omega$;  $m$ and $n$ are the gravitational mass and the magnetic mass or NUT parameter, respectively; $\epsilon$ and $k$ are kinematic parameters that can be fixed as $\epsilon=-1,0,1$ and $k\in \mathbb N$. \newline
This metric obeys the Einstein-Maxwell equations with an electromagnetic vector potential given by
\begin{equation*}
    \textbf{A}= - \frac{1}{\Sigma}[eq(\textbf{d}\tau-\omega p^2\textbf{d}\phi) + gp(\omega\textbf{d}\tau+q^2\textbf{d}\phi) ].
\end{equation*}
Additionally, the Pleba{\'n}ski metric possesses the Killing-Yano tensor \cite{Car87}: 
\begin{equation}
    \textbf{k} =[\omega p\textbf{d}q\wedge (\textbf{d}\tau-\omega p^2\textbf{d}\phi) + q\textbf{d}p\wedge (\omega\textbf{d}\tau+q^2\textbf{d}\phi) ]\;.
\end{equation}
Is is known that the dual $h$,  of a Killing-Yano tensor $k$, $\textbf{h}=*\textbf{k}$ is also a Killing-Yano tensor,
\begin{equation}
    \textbf{h} = [q\textbf{d}q\wedge(\omega p^2\textbf{d}\phi -\textbf{d}\tau) + \omega p\textbf{d}p\wedge(q^2\textbf{d}\phi+\omega\textbf{d}\tau) ],
\end{equation}
or $\textbf{h}=\textbf{db}$, where $2\textbf{b}=(p^2-q^2)\textbf{d}\tau+p^2q^2\textbf{d}\phi$. We note that neither $\textbf{k}$ nor $\textbf{h}$ depend on the metric functions P(p) and Q(q), this fact will be exploited later.
\newline
From the two Killing-Yano tensors we can construct three tensors: one related to $\textbf{k}$ as $(K_k)_{\alpha \beta} =  \textbf{k}_{\alpha \rho} \textbf{k}_\beta^{\;\;\rho}$; the second related to $\textbf{h}$,  $(K_h)_{\alpha \beta} =  \textbf{h}_{\alpha \rho} \textbf{h}_\beta^{\;\;\rho}$;  and a third related to both by a symmetrized product $(K_{kh})_{\alpha\beta}= \frac{1}{2}(\textbf{k}_{\alpha\rho} \textbf{h}^{\;\;\rho} _\beta + \textbf{h}_{\alpha\rho} \textbf{k}^{\;\;\rho} _\beta)$. Such that the resulting tensors associated to the Pleba{\'n}ski metric are:
\begin{equation}
  \begin{split}
      K_k & = \left[ \frac{Q\omega^2p^2 (\textbf{d}\tau-\omega p^2\textbf{d}\phi)^2}{\Sigma} + \frac{Pq^2 (\omega \textbf{d} \tau +q^2\textbf{d}\phi)^2}{\Sigma} + \frac{\Sigma q^2} {P} \textbf{d} p^2 - \frac{\Sigma \omega^2p^2}{Q} \textbf{d} q^2 \right]\;,\\
      K_h & = \left[ \frac{Q q^2 (\textbf{d}\tau-\omega p^2\textbf{d}\phi)^2}{\Sigma} + \frac{P\omega^2p^2 (\omega \textbf{d} \tau +q^2\textbf{d}\phi)^2}{\Sigma} + \frac{\Sigma \omega^2p^2} {P} \textbf{d} p^2 - \frac{\Sigma q^2}{Q} \textbf{d} q^2 \right]\;,\\
      K_{kh} & =  p q \omega\left[ -\frac{Q(\textbf{d}\tau-\omega p^2\textbf{d}\phi)^2}{\Sigma} + \frac{P(\omega \textbf{d}\tau+q^2\textbf{d}\phi)^2}{\Sigma} + \frac{\Sigma}{P}\textbf{d}p^2 + \frac{\Sigma}{Q}\textbf{d}q^2\right]\;.
  \end{split}
  \label{KTs}
\end{equation}
Note that the tensor $K_{k h}$ is proportional to the metric $\textbf{g}$, Eq. (\ref{PD}), and the others are related to $\textbf{g}$ by
\begin{equation*}
    (\omega^2p^2-q^2)\textbf{g} = (\omega^2p^2-q^2)/pq\omega \;K_{kh} =  K_h-K_k.
\end{equation*}
We focus on the $K_k$ tensor;  it fulfills the Killing tensor equation, $(K_k)_{(\alpha \beta;\gamma)}=0$, so that there is   associated one conserved quantity, $C$, that can be computed according to $K_{\alpha \beta} \Dot{x}^\alpha \Dot{x}^\beta = C$, where $\Dot{x}^\beta$ are the 4-velocities of a test particle.
\newline

To implement the generalization,  we perform a deformation of the Pleba{\'n}ski metric functions. Let us consider the co-tetrad
\begin{equation*}
    e^0 = \left(\frac{Q}{\Sigma} \right)^{1/2}(d\tau-\omega p^2d\phi),\;\; e^1 = \left(\frac{P}{\Sigma} \right)^{1/2}(\omega d \tau +q^2d\phi),\;\; e^3=\left(\frac{\Sigma}{P} \right)^{1/2}dp,\;\; e^4=\left(\frac{\Sigma}{Q} \right)^{1/2}dq.
\end{equation*} 
Introducing the functions $\Xi(q)$, $\Phi(q)$ substituting in the metric   
\begin{equation}
    q^2 \longrightarrow (\Xi(q))^2; \;\;\;\; e^0 \longrightarrow e^{-\Phi(q)} e^0; \;\;\;\; \Sigma \longrightarrow \Sigma = \Xi^2 + \omega^2p^2,
\end{equation}
and allowing  that $P(p)$ and $Q(q)$ be general but well behaved functions, we have 

\begin{equation}
   ds^2= -e^{-2\Phi(q)}\frac{Q(\textbf{d}\tau-\omega p^2\textbf{d}\phi)^2}{\Sigma} + \frac{P(\omega\textbf{d}\tau+ \Xi(q)^2\textbf{d}\phi)^2}{\Sigma} + \frac{\Sigma}{P}\textbf{d}p^2 + \frac{\Sigma}{Q}\textbf{d}q^2.
\label{PD4}
\end{equation}
Note that, when $\Xi=q$ and $\Phi=0$ we recover the Pleba{\'n}ski metric. In order to preserve the Pleba{\'n}ski asymptotics we impose the following conditions on the metric functions, 
\begin{equation}
    Q(q) \thicksim q^4; \;\;\;\; \Phi \thicksim o(1);\;\;\;\; \Xi(q) \thicksim q; \;\;\;\; P(p) \thicksim p^4;
\end{equation}
with these changes, we depart from the Ricci-flat spacetime, with
Ricci scalar $R= - 4 \Lambda$ derived by Pleba\'nski in \cite{Pleb}, where it is assumed that $P'' + Q'' = -4\lambda \Sigma$, with $\lambda$ being a separation constant. Moreover, this metric is no longer of Petrov type D, as all Weyl invariants are non-null, it turns out that this metric is Petrov type I or algebraically general. See Appendix B for the algebraic classification of the metric and the definition of the curvature invariants.
\newline
In the non-rotating case, $\omega \xrightarrow{} 0$ we recover the Reissner-Nordstrom-like metric with generalized functions. For the co-tetrad, $\Sigma \rightarrow \Xi(q)^2$
\begin{equation}
    e^0 = e^{-\Phi(q)}\frac{Q^{1/2}}{\Xi(q)} \textbf{d}\tau,\;\; e^1 = P^{1/2}\Xi(q)\textbf{d}\phi,\;\; e^3= \frac{\Xi(q)}{P^{1/2}}\textbf{d}p, \;\; e^4 = \frac{\Xi(q)}{Q^{1/2}} \textbf{d}q;
\end{equation}
while for the line element
\begin{equation}
   ds^2= -e^{-2\Phi(q)}\frac{Q}{\Xi(q)^2}\textbf{d}\tau^2 + \frac{\Xi(q)^2}{Q}\textbf{d}q^2 + \Xi(q)^2\left[P\textbf{d}\phi^2 + \frac{1}{P}\textbf{d}p^2 \right].
   \label{spherical}
\end{equation}
The static case will be examined in section 5, where some examples of sources for this geometry will be presented. Another important particular case of the Pleba{\'n}ski spacetime, is obtained via the transformation
\begin{equation}
    \tau \xrightarrow{} \tau-\omega\phi,\;\;\; \phi \xrightarrow{}- \phi, \;\;\; p \xrightarrow{} \cos\theta, \;\;\; q \xrightarrow{} r\;.
\end{equation}
that leads to the line element
\begin{equation}
     ds^2 = -e^{-2\Phi(r)}\frac{Q}{\Sigma}(\textbf{d}\tau - \omega \sin^2\theta\;\textbf{d}\phi)^2 + \frac{P}{\Sigma}(\omega\textbf{d}\tau - (\omega^2 + \Xi(r)^2)\textbf{d}\phi)^2 + \frac{\Sigma} {P} \sin^2\theta\;\textbf{d}\theta^2 + \frac{\Sigma}{Q}\textbf{d}r^2,
\end{equation}
that is a generalization of the Kerr metric. It can be asymptotically ``Kerr-like'' spacetime by imposing conditions $Q(r) \thicksim r^2; \; \Phi \thicksim o(1);\; \Xi(r) \thicksim r; \;P(\cos\theta) \thicksim \cos^2\theta$.
\newline

Finally, these transformations cast  the electromagnetic potential as
\begin{equation}
    \textbf{A}= - \frac{1}{\Sigma}[e\Xi(q)e^{-\Phi(q)}(\textbf{d}\tau-\omega p^2\textbf{d}\phi) + gp(\omega\textbf{d}\tau+\Xi(q)^2\textbf{d}\phi) ].
\label{A}
\end{equation}
While the Faraday tensor, $\textbf{F}=\textbf{dA}$, is given by
\begin{equation*}
 \begin{split}
    \Sigma^2\textbf{F} 
        & =[ e (2\Xi^2\Xi'-\Sigma\Xi' + \Sigma\Xi\Phi') + 2g\Xi\Xi'\omega p e^{\Phi} ] \textbf{d}q \wedge \textbf{M} + [ 2e\Xi\omega p e^{-\Phi} + g(\omega^2p^2-\Xi^2 ) ] \textbf{d}p \wedge \textbf{N}. 
 \end{split}
\end{equation*}
Where the one-forms, $\textbf{M}$ and $\textbf{N}$ are, $\textbf{M}=e^{-\Phi}(\textbf{d}\tau-\omega p^2\textbf{d}\phi)$ and  $\textbf{N}=\omega\textbf{d}\tau+\Xi^2\textbf{d}\phi$.
The components of the Faraday tensor are given by
\begin{equation*}
 \begin{split}
          F_{q \tau} &= \frac{e^{-\Phi}}{\Sigma^2} \left[ 2ge^{\Phi}\Xi'\Xi \omega p - e(\Xi^3 \Phi' + (\Xi \Phi' - \Xi') \omega^2p^2 + \Xi'\Xi^2)  \right] \\
         F_{p \tau} &= - \frac{\omega}{\Sigma^2} \left[ g(\Xi^2-\omega^2p^2) - 2e\Xi\omega p e^{-\Phi}  \right],\\
         F_{p \phi} & = \frac{\Xi^2}{\omega} F_{p \tau}, \; \quad  F_{q \phi} = -\omega p^2 F_{q \tau}
 \end{split}
\end{equation*}
thus, there are only two independent components.
It is straightforward but unexciting to verify that $\textbf{A}$ is an exact  closed form, as $\textbf{dF}=\textbf{d}^2\textbf{A} = 0 $. 
It turns out that the spacetime (\ref{PD4}) is no longer compatible with Maxwell's equation. However,  the metric (\ref{PD4}) can be sourced by nonlinear electromagnetic fields,  as we show in Sect. \ref{sect5}.

In what follows, the separability properties of the general metric (\ref{PD4})  are analyzed and the conditions required on the general functions to guarantee separability are determined.
\subsection{Hamilton-Jacobi separability of the generalized Pleba{\'n}ski metric}

An important aspect of the separability of the HJ equation is that it excludes the possibility of chaotic particle orbits, which agrees with the observation of stable accretion disks in the vicinity of BHs.

In a  metric $g_{\mu \nu}$ the geodesic motion equations can be obtained by solving the Hamilton-Jacobi (HJ) equation
\begin{equation}
    \frac{\partial S}{\partial \lambda} + g^{\mu\nu}\partial_\mu S\partial_\nu S = 0,
\label{HJeq}
\end{equation}
where $\lambda$ is an affine parameter of the geodesic motion. The HJ equation is separable in the spacetime $g_{\mu \nu}$ if Eq. (\ref{HJeq}) can be separated with the ansatz that the solution is 
a sum of functions $F_\mu$ that only depend on the $x^\mu$ coordinate and on the constants of motion $C_a$,
\begin{equation*}
    S = \sum_{\mu=0}^3 F_\mu(x^\mu, C_a),
\end{equation*}
where $a=0, 1,2,3$, since  for a full separable HJ equation four constants of motion are needed.
In a stationary axisymmetric metric there are two motion constants associated with the two Killing vectors, $\partial_{\tau}$ and $\partial_{\phi}$; the mass of the test particle is a third motion constant, and in case the metric possesses one Killing tensor  there is a fourth motion constant that makes the system completely integrable.

For a stationary axisymmetric spacetime the HJ equation is separable if the contravariant  metric, in $(\tau, \phi, p, q)$ coordinates, 
has the block structure \cite{Btt}:
\begin{equation}
 g^{\mu \nu}= (A_1 + B_1)^{-1}
\begin{pmatrix}
A_5 + B_5 &   A_4 + B_4 &   0   &   0\\
A_4 + B_4 &   A_3 + B_3 &   0   &   0\\
    0     &       0     &  A_2  &   0\\
    0     &       0     &   0   &  B_2\\
\label{mg}
\end{pmatrix}
\end{equation}
where $A_i=A(p)$ and $B_i=B(q)$, $i=1,...,5$ are ten arbitrary and well-behaved functions. In the generalized PD spacetime case (\ref{PD4}) the functions $A_i(p)$ and $B_i(q)$ are
\begin{equation*}
 \begin{split}
     A_1 & = \omega^2p^2, \;\;\; A_2 = P, \;\;\; A_3 = 1/P, \;\;\; A_4 =\omega p^2/P, \;\;\; A_5 = \omega^2p^4 /P\\
     B_1 & = \Xi(q)^2, \;\; B_2 = Q, \;\;\; B_3 = -\omega^2 / Qe^{-2\Phi}, \;\;\; B_4 =\omega \Xi(q)^2/Qe^{-2\Phi}, \;\;\; B_5 = -\Xi(q)^4 /Qe^{-2\Phi}.
 \end{split}
\end{equation*}
Note that the functions $A_3, A_4, A_5$ can be expressed in terms of $A_1$ and $A_2$, meaning that only the function $P(p)$ appears as a general factor. Additionally, the functions $B_4$ and $B_5$ can also be written in terms of  $B_1, B_2, B_3$, leaving  only a possible three-function generalization. Even an additional function on the angular coordinate $p$ can be introduced by transforming $\omega  p^2 \rightarrow \Theta(p)$, but this does not provide any additional advantage, in contrast to the case of the coordinate $q$ is related to the radial distance. 
\newline
In this context, the determinant of the metric is $g=det(g_{\mu \nu})=- e^{-2 \Phi}(A_1 + B_1)^2$; again we remark that it does not depend on the functions $P(p)$ and $Q(q)$. The Killing tensor $K$ is given by
\begin{equation}
 K^{\mu \nu}= (A_1 + B_1)^{-1}
\begin{pmatrix}
B_1A_5 - A_1B_5 &  B_1A_4 - A_1B_4&   0   &   0\\
B_1A_4 - A_1B_4 &  B_1A_3 - A_1B_3&   0   &   0\\
       0        &         0       & A_2B_1&   0\\
       0        &         0       &   0   &-B_2A_1\\
\end{pmatrix}
\end{equation}
From this tensor, a conserved quantity $C$, can be derived by contracting it with the 4-velocities of the test particle,  $K_{\alpha \beta} \Dot{x}^\alpha \Dot{x}^\beta = C$. This fourth constant allows the determination of the first integrals, $\dot{p}$ and $\dot{q}$; for the Kerr metric $C$ is known as the Carter constant \cite{Car68}. 
To illustrate the advantage of symmetries in the integration of the motion equations, in the next subsection, the trajectories of uncharged and charged particles are determined for the deformed or generalized Pleba{\'n}ski spacetimes. Motion constants enable us to integrate the 4-velocities of a test particle just through algebraic methods.
\subsubsection{Uncharged test particle trajectories in the deformed Pleba{\'n}ski metric}
The  4-velocity vector of the test particle is $u^\mu=dx^\mu/d\lambda = \dot{x}_\mu$, with $\lambda$ being an affine parameter, which will henceforth be represented in the form $u^\mu = (\Dot{\tau},\Dot{\phi}, \Dot{p},\Dot{q})$. 
We will exploit the advantage of having two Killing vectors for determining two components of the 4-velocity. There is a covector $u_\mu = (c_\tau,c_\phi, u_p,u_q)$, where $c_\tau$ and $c_\phi$ are the constants related to the symmetries in $\tau$ and $\phi$, respectively.  
\newline
Given $u_\mu=g_{\mu \nu}u^\nu$, we can find $\Dot{\tau}$ and $\Dot{\phi}$ in terms of $c_\tau$, $c_\phi$ and the metric functions, such as

\begin{equation}
\begin{split}
    \Dot{\tau} & =  \frac{c_\tau g_{\phi \phi} - c_\phi g_{\tau \phi}}{g_{\tau\tau}g_{\phi\phi}-(g_{\tau \phi})^2}\;,\\
    \Dot{\phi} & = -\frac{c_\tau g_{\tau \phi} - c_\phi g_{\tau \tau}}{g_{\tau\tau}g_{\phi\phi}-(g_{\tau \phi})^2}\;.
\end{split}
\label{sol TF}
\end{equation}
The components $u_{p}$ and $u_{q}$ are determined  using the motion constants $\mu$ and $C$ generated by the Killing tensors $K$ and $g$,
\begin{equation}
\begin{split}
g_{\mu \nu}\dot{x}^\mu\dot{x}^\nu &= g^{\mu \nu}\dot{x}_\mu\dot{x}_\nu = - \mu^2, \\
K_{\mu \nu}\dot{x}^\mu\dot{x}^\nu &= K^{\mu \nu}\dot{x}_\mu\dot{x}_\nu = C;
\label{sisgk}
\end{split}
\end{equation}
then $u_{p}$ and $u_{q}$  are
\begin{equation}
 \begin{split}
    u_p^2 &= -\frac{A_1}{A_2}\mu^2 + \frac{C}{A_2} - \frac{1}{A_2}\left[c_{\tau}^2 A_5 + 2c_\tau c_\phi A_4 +c_{\phi}^2 A_3 \right],\\
    u_q^2 & = -\frac{B_1}{B_2}\mu^2 - \frac{C}{B_2} - \frac{1}{B_2}\left[c_{\tau}^2 B_5 + 2c_\tau c_\phi B_4 +c_{\phi}^2 B_3 \right].
 \end{split}
 \label{upq}
\end{equation}
Substituting $A_i$ and $B_i$  we arrive at $\dot{p}$ and $\dot{q}$,
\begin{equation}
\begin{split}
\dot{p}^2 &
 = -\frac{P\omega^2p^2}{\Sigma^2}\mu^2 + \frac{PC}{\Sigma^2} - \frac{\left(c_{\tau}\omega p^2 +c_{\phi} \right)^2}{\Sigma^2}, \\
    \dot{q}^2 & =  -\frac{Q \Xi^2(q)}{\Sigma^2}\mu^2 - \frac{QC}{\Sigma^2} + \frac{\left(c_{\tau}\Xi^2(q)- c_{\phi} \omega \right)^2}{e^{-2\Phi}\Sigma^2}.
 \end{split}
\end{equation}
These velocities can also be derived from the Hamilton Jacobi equations \cite{Pleb}, but the advantage of using the motion constants is the simplicity of the computations,  i.e. we do not need to do integrals. The presence of $\Sigma$ in  $\dot{p}$ and $\dot{q}$ indicates a coupling  on $q$ and  $p$; in some works this question has been solved introducing the so called Mino time that decouples the radial and polar motion of a particle \cite{Minos}.
\subsubsection{Trajectories of a charged particle in the deformed Pleba{\'n}ski spacetime}
If we consider a test electric charge $e_0$, its movement is governed by
\begin{equation}
 \left(\frac{d}{d\lambda}\Dot{x}^\mu + \Gamma ^\mu _{\alpha \beta} \Dot{x}^\alpha \Dot{x}^\beta \right) = - e_0 F_\nu ^{\; \; \mu} \Dot{x}^\nu.
\label{tray}
\end{equation}
Again we can solve the equations for $\Dot{\tau}$ and $\Dot{\phi}$; and then using the contractions $g_{\mu \nu} \Dot{x}^\mu \Dot{x}^\nu =  -\mu^2$ and $K_{\mu \nu} \Dot{x}^\mu \Dot{x}^\nu = C$ we can determine $\Dot{p}$ and $\Dot{q}$. Equivalently,  we can use the Hamilton-Jacobi formalism since the metric is separable \cite{Pleb}; both methods are equivalent to solving Eq. (\ref{tray}). \newline
The components $\Dot{\tau}$ and $\Dot{\phi}$ are obtained  by making $c_\tau \xrightarrow{} c_{\tau} - e_0\textbf{A}_\tau$ and $c_\phi \xrightarrow{} c_{\phi} - e_0\textbf{A}_\phi$ in Eq. (\ref{sol TF}), where $\textbf{A}_\tau$ and $\textbf{A}_\phi$ are the components of the electromagnetic 4-potential (\ref{A}),
\begin{equation}
\begin{split}
      \Dot{\tau} & = \Dot{\tau_0} -e_0\frac{ P \Xi^2(\omega \textbf{A}_\phi - \Xi^2 \textbf{A}_\tau) + Qp^2\omega(\textbf{A}_\phi +\omega  p^2 \textbf{A}_\phi) }{ PQ\Sigma } = \Dot{\tau}_0 + u\;,\\
    \Dot{\phi} & = \Dot{\phi_0} - e_0\frac{P\omega(\Xi^2\textbf{A}_\tau-\omega \textbf{A}_\phi) + Q(\omega  p^2 \textbf{A}_\tau + \textbf{A}_\phi) }{PQ\Sigma} = \Dot{\phi}_0 + v\;,
\end{split}
\label{tpeg}
\end{equation}
the functions $\Dot{\tau_0}$, $\Dot{\phi_0}$ are the same than in Eq. (\ref{sol TF}) and the functions $u$ and $v$, substituting $A_{\mu}$, are given by:
\begin{equation*}
     u = - e_0 \frac{Pe \Xi(q)^3e^{\Phi}-Q g\omega  p^3} {PQ\Sigma} ,\;\;\; v=e_0 \frac{ P e\omega\Xi e^{\Phi} + Q gp} {PQ\Sigma}\;.
\end{equation*}
\newline
The contractions of the metric and the St{\"a}ckel-Killing tensor with the 4-velocities are given in Eq. (\ref{sisgk}) and solving for $\Dot{p}$ and $\Dot{q}$ we obtain,
\begin{equation}
 \begin{split}
    \dot{p}^2 &= -\frac{P\omega^2p^2}{\Sigma^2}\mu^2 + \frac{PC}{\Sigma^2} - \frac{\left(c_{\tau}\omega  p^2 + ge_0p +c_{\phi} \right)^2}{\Sigma^2}, \\
    \dot{q}^2 &= -\frac{Q \Xi(q)^2}{\Sigma^2}\mu^2 - \frac{QC}{\Sigma^2} + \frac{\left(c_{\tau}\Xi(q)^2 + ee_0\Xi(q)e^{-\Phi}- c_{\phi} \omega \right)^2}{e^{-2\Phi}\Sigma^2}.
 \end{split}
\end{equation}
This method of determining $\Dot{p}$ and $\Dot{q}$ is faster than by using the Hamilton-Jacobi equation or solving the Lorentz force equation (\ref{tray}),  as  presented in \cite{Pleb}.  Despite having an associated Killing tensor and therefore admitting the separability of the HJ equation, it does not imply the separability of the generalized Pleba{\'n}ski metric for other equations, such as the wave equation \cite{PpKt}. 
\subsection{Klein-Gordon separability in the deformed Pleba{\'n}ski spacetime}
On a curved background $g_{\mu\nu}$ the dynamics of a massive scalar field $\Phi$ is governed by the Klein-Gordon (KG) equation
\begin{equation}
    \Box\Phi = \frac{1}{\sqrt{|g|}}\partial_\mu(\sqrt{|g|} g_ {\mu\nu} \partial_\nu\Phi) = m^2 \Phi.
    \label{KGeq}
\end{equation}
Consider that the metric $g_{\mu\nu}$ is cyclic in the coordinates $x^j$, $j \le 4$, while the remaining coordinates $l=4-j$ $x^l$ appear explicitly in the metric; then  the ansatz  \cite{Mill}
\begin{equation*}
\Phi(x^\mu)=\prod_j e^{i C_j x^j}\prod_{l=4-j}^4\Psi_l(x^l),
\end{equation*}
leads to separability,
where the sum over $j$ is assumed and  $C_j$ are constants.
\newline
On the other hand, separability of the KG equation is guaranteed if the anomaly-free condition  is satisfied \cite{Car77},
\begin{equation}
    \nabla_\alpha[K,R]^\alpha_{\;\beta} = \nabla_\alpha(K^\alpha_{\;\;\sigma} R^\sigma_{\;\;\beta} - R^\alpha_{\;\;\sigma} K^\sigma_{\;\;\beta}) = 0.
\label{KR}
\end{equation}
In this subsection,  we shall prove that the Ricci and Killing tensors of   generalized Pleba{\'n}ski metric fulfill  the anomaly-free condition.

Let us consider the ``mixed'' components of the Ricci tensor; the indices  $i,j,k$ run over the cyclic coordinates  $\tau, \phi$ and $l,m,n=p,q$,  to distinguish between the conserved quantities and the rest,
\begin{equation*}
 \begin{split}
      R_{i l} & = - \Gamma^\mu_{\;\;i\mu,l} + \Gamma^\mu_{\;\;il,\mu} + \Gamma^\mu_{\lambda\mu}\Gamma^\lambda_{il} - \Gamma^\mu_{\lambda l}\Gamma^\lambda_{i\mu} \\
      & = \Gamma^j_{l j}\Gamma^l_{il} + \Gamma^m_{l m}\Gamma^l_{il} - \Gamma^j_{k l}\Gamma^k_{ij} - \Gamma^m_{n l}\Gamma^n_{im} = 0.
 \end{split} 
\end{equation*}
Thus, both $R$ and $K$, and their products  $R^{\alpha \rho} K_{\rho \beta}$ and $K^{\alpha \rho} S_{\rho \beta}$ have the form
\begin{equation*}
\left[M_{\mu \nu} \right] = \left(
\begin{matrix}
A & 0_{2 \times 2} \\
0_{2 \times 2} & B \\
\end{matrix}
\right),
\end{equation*}
where $A$, $B$ are non null matrices $2 \times 2$. Denoting  $S^\alpha_\beta =[K,R]^\alpha_{\;\beta}$, the trace,  
\begin{equation*}
    S^\alpha_{\;\alpha} =[K,R]^\alpha_{\;\alpha} = (K^\alpha_{\;\;\sigma} R^\sigma_{\;\;\alpha} - R^\alpha_{\;\;\sigma} K^\sigma_{\;\;\alpha}) = 0.
\end{equation*}
If $\alpha=i$, $\beta=l$, note that the Killing tensors (\ref{KTs}) fulfill $K^l_{\;m}=0$ if $l\not=m$
\begin{equation*}
    S^i_l =K^i_{\;\;\sigma} R^\sigma_{\;\;l} - R^i_{\;\;\sigma} K^\sigma_{\;\;l} = K^i_{\;\;j} R^j_{\;\;l} - R^i_{\;\;l} K^l_{\;\;l} = 0;
\end{equation*}
and if $\alpha=l$, $\beta=i$,
\begin{equation*}
    S^l_i =K^l_{\;\;\sigma} R^\sigma_{\;\;i} - R^l_{\;\;\sigma} K^\sigma_{\;\;i} = K^l_{\;\;l} R^l_{\;\;i} - R^l_{\;\;j} K^j_{\;\;i} = 0.
\end{equation*}
To find $\nabla_\alpha S^\alpha_\beta$, explicitly
\begin{equation*}
\nabla_\alpha S^\alpha_\beta = \partial_\alpha S^\alpha_\beta + \Gamma^\alpha_{\alpha\mu}S^\mu_\beta - \Gamma^\mu_{\alpha\beta} S^\alpha_\mu;
\end{equation*}
let it  be $\beta=i$
\begin{equation*}
\begin{split}
\nabla_\alpha S^\alpha_i & =  \partial_\alpha S^\alpha_i + \Gamma^\alpha_{\alpha\mu}S^\mu_i - \Gamma^\mu_{\alpha i} S^\alpha_\mu = \Gamma^\alpha_{\alpha j}S^j_i - \Gamma^\mu_{j i} S^j_\mu - \Gamma^\mu_{l i} S^l_\mu\\
     &= (\Gamma^k_{k j}+ \Gamma^l_{l j})S^j_i - \Gamma^k_{j i} S^j_k - \Gamma^m_{l i} S^l_m  = 0.
 \end{split}
\end{equation*}
 Considering that
 \begin{equation*}
  \begin{split}
    S^m_{\;l} & = [K,R]^m_{\;l} = K^{m\sigma}R_{\sigma l} - R^{m\sigma}K_{\sigma l} = K^{m m}R_{m l} - R^{m l}K_{l l}  \\
    & = K^{m m}R_{m l} - g^{m \sigma} R_{\sigma l} K^{l \rho} g_{\rho l}= \left( K^{m m} - g^{m m } g_{l l} K^{l l}  \right) R_{m l},
  \end{split}
 \end{equation*}
 in the last equation the sum convention does not apply, the indexes $m,l$ are fixed indicating the trace elements; thus
 \begin{equation*}
S^p_{\;q} = \left( K^{pp} - g^{pp} g_{qq} K^{qq}  \right) R_{pq} = \left( \frac{P \Xi^2}{\Sigma}  + \frac{\Sigma}{Q} \frac{P}{\Sigma} \frac{Q\omega^2 p^2}{\Sigma} \right) R_{pq} = P R_{pq},
\end{equation*}
and
\begin{equation*}
     S^q_{\;p} = \left( K^{qq} - g^{qq} g_{pp} K^{pp}  \right) R_{pq} = \left(-\frac{Q\omega^2 p^2}{\Sigma} - \frac{\Sigma}{P} \frac{Q}{\Sigma}\frac{P \Xi^2}{\Sigma} \right) R_{pq}  = -Q R_{pq},
 \end{equation*}
but $R_{pq} = 0$. This shows that Eq.(\ref{KR}) is satisfied. The KG separability condition (\ref{KR}) can also be easily tested with a specialized program like GRTensor; we have used a more traditional approach using the shape of the  involved tensors to prove the KG separability in the deformed Pleba{\'n}ski metric.
Therefore,  the conclusion of this section is that the generalized metric functions preserve the symmetries related to the separability of both, the HJ and KG equations, which implies there are no chaotic trajectories of test particles.

\section{Killing horizons and surface gravity of the generalized Pleba{\'n}ski metric}
\label{sect3}
One of the important properties of stationary axisymmetric metrics is the existence of horizons and then their relationship with BHs.
If we have a Killing vector $k$ describing a stationary spacetime, then at certain region $k$ may become lightlike, i.e. $k_\mu k^\mu = 0$; if this region forms a hypersurface $\Sigma$, then this null hypersurface is called a Killing horizon.

In the thermodynamics associated to a BH a key role is played by surface gravity that for a general black hole is not well
defined. However, it can be defined if  the BH event horizon is a Killing horizon. The surface gravity $\kappa$ of a static Killing horizon is the acceleration, exerted at infinity, needed to keep an object on the event horizon. 

We check that the event horizon of the generalized Pleba\'nski metric is a Killing horizon and then obtain the expression for the surface gravity.
 
The location of the horizons in the Pleba{\'n}ski deformed spacetime is determined by the zeros of  the metric component $g^{qq}$.  Labelling all the zeros of the  function $Q(q)$ and checking that $\dot{q}=cte$, so that the Killing vector is confined to the surface delimited by the horizon. Then a general Killing vector is constructed, and the parameter that bounds the vector to the horizon is determined,
\begin{equation*}
    \left\{ q_{H_i}: Q(q_{H_i}) \right\},
\end{equation*}
where $H_i$ denotes the $i$-horizon since at these points
\begin{equation*}
    g^{\mu \nu}\partial_\mu q \partial_\nu q = g^{qq} = \frac{Q(q)}{\Sigma(p,q)} \rightarrow 0, 
\end{equation*}
that defines a null surface. 
To ensure that these are true event horizons, the geodesics along  the $\tau$ or $\phi$ direction should stay on the surface defined by the horizon; this is
\begin{equation*}
\frac{d^2q}{d\lambda^2} = - \left[\Gamma^q_{a b} \frac{dx^a}{d \lambda} \frac{dx^b}{d \lambda} \right]_{q=q_{H_i}} = 0,
\end{equation*}
where $a,b=\tau, \phi, p$. We check that $\Gamma$ vanishes at the horizon,
\begin{equation*}
\Gamma^q_{a b} = \frac{g^{qq}}{2}(g_{q(a,b)} - g_{ab,r}) = -\frac{g^{qq}}{2} g_{ab,r}  \propto Q (q),
\end{equation*}
so $\Gamma^q_{a b} \rightarrow 0$ on the horizons. \newline
To prove that these are indeed Killing horizons, we need to prove the existence of a Killing vector whose norm is null when evaluated at the horizons. The Pleba{\'n}ski metric possesses two Killing vectors, one associated to time translation, $\xi^\mu$, and the second one to axial symmetry, $\chi^\mu $,
\begin{equation*}
    \xi^\mu = (1,0,0,0); \;\; \chi^\mu = (0,1,0,0);
\end{equation*}
and we construct a general Killing vector as a linear combination of them
\begin{equation*}
    \kappa^\mu = (1,\Lambda,0,0),
\end{equation*}
where $\Lambda$ is a constant.  We wish to find a collection of constants $\Lambda_{H_i}$ such that the norm of $\kappa^\mu\kappa_\mu$ vanishes on the respective horizon $q_{H_i}$. To that end it is enough to impose that
\begin{equation*}
\left[g_{\mu\nu}\kappa^\mu \kappa^\nu\right]_{q=q_{H_i}} = \left[g_{\tau \tau} + 2\Lambda g_{\tau \phi} + \Lambda^2g_{\phi \phi} \right]_{q=q_{H_i}} = 0,
\end{equation*}
substituting the metric functions yields
\begin{equation*}
\left[g_{\mu\nu}\kappa^\mu \kappa^\nu\right] = \frac{P}{\Sigma} \left[\omega+ \Lambda\Xi^2\right]^2 - \frac{Qe^{-2\Phi}}{\Sigma} \left[1 + \Lambda\omega p^2\right]^2,
\end{equation*}
so that
\begin{equation*}
\left[g_{\mu\nu}\kappa^\mu \kappa^\nu\right]_{q=q_{H_i}} = \frac{P \left[\omega+ \Lambda\Xi^2(q_{H_i})\right]^2}{\Xi^2(q_{H_i}) + \omega^2 p^2};
\end{equation*}
in order to have a Killing horizon we demand
\begin{equation*}
\Lambda_{H_i} = -\frac{\omega}{\Xi^2(q_{H_i})}.
\end{equation*}
Such that the horizon is a Killing one.

\subsection{Surface Gravity of the generalized Pleba{\'n}ski  metric}
To determine the surface gravity for the generalized Pleba{\'n}ski, we first calculate the area of the surfaces of constant $q$, given by
\begin{equation*}
\begin{split}
S(q)  & = 2\pi \int_{-1}^{1} (g_{\phi \phi} g_{pp})^{1/2} dp \\
& = 2\pi \int_{-1}^{1} (\Xi^4-Qe^{-2\Phi}\omega^2p^4/P)^{1/2} dp;
\end{split}
\end{equation*}
that at the horizon  simplifies,
\begin{equation*}
S(q_{H_i}) = 2\pi \int_{-1}^{1} \Xi^2 dp 
= 4\pi \Xi^2(q_{H_i}).
\end{equation*}
Taking the limit case $\Xi^2(q) \rightarrow q^2$ we recover the area of the sphere, as expected. 

The surface gravity for a Killing horizon is given by \cite{Wald}
\begin{equation*}
 |\kappa| = \sqrt{\frac{1}{2}(\nabla_\mu\kappa^\nu)(\nabla_\nu\kappa^\mu)}\biggm|_H, 
\end{equation*}
where
\begin{equation*}
    \nabla_\mu\kappa^\nu = \kappa^\nu_{\;,\mu} + \Gamma^\nu_{\mu \sigma} \kappa^\sigma = \Gamma^\nu_{\mu i} \kappa^i,
\end{equation*}
thus
\begin{equation*}
 \begin{split}
       |\kappa| & = \sqrt{\frac{1}{2}(\Gamma^\nu_{\mu i} \kappa^i)(\Gamma^\mu_{\nu j} \kappa^j)}\biggm|_H = \sqrt{\frac{1}{2}(\Gamma^\nu_{k i}\Gamma^k_{\nu j} + \Gamma^\nu_{l i}\Gamma^l_{\nu j} )\kappa^i \kappa^j}\biggm|_H \\
       & =\sqrt{\frac{1}{2}(\Gamma^l_{k i}\Gamma^k_{l j} + \Gamma^k_{l i}\Gamma^l_{k j} )\kappa^i \kappa^j}\biggm|_H = \sqrt{\Gamma^k_{l i}\Gamma^l_{k j}\kappa^i \kappa^j}\biggm|_H.
 \end{split}
\end{equation*}
Evaluating on the horizons $q=q_{H_i}$ means that $Q=0$ and $\Lambda = -
\omega/\Xi^2$,
\begin{equation*}
    |\kappa| = \sqrt{\frac{(\Lambda\omega p^2-1)^2}{4\Sigma^2} e^{-2\Phi} Q'^2 }\biggm|_H = \frac{e^{-\Phi(q_{H_i})} Q'(q_{H_i})}{2 \Xi(q_{H_i}) }.
\end{equation*}
This  expression generalizes surface gravity  in \cite{BaVis}. 
\section{Conformal factor in the Pleba{\'n}ski metric}\label{sect4}
In the attempt to obtain regular solutions to the Einstein equations, including a conformal factor is an admissible procedure to remove singularities, and appropriate choices for the conformal factor can lead to geodesically complete spacetimes \cite{Fran}. 

Therefore,  the metric functions generalization of the Pleba{\'n}ski metric is enhanced by introducing a conformal factor $\Omega$. However,  there is an impact on the separability of the HJ and KG equations. Let us consider the conformal metric $\bar{g}_{\mu\nu} \rightarrow \Omega^2 g_{\mu\nu} $ and its contravariant  $\bar{g}^{\mu\nu} = \Omega^{-2} g^{\mu\nu} $,
\begin{equation}
 \bar{g}^{\mu \nu}= \Omega^{-2}(A_1 + B_1)^{-1}
\begin{pmatrix}
A_5 + B_5 &   A_4 + B_4 &   0   &   0\\
A_4 + B_4 &   A_3 + B_3 &   0   &   0\\
    0     &       0     &  A_2  &   0\\
    0     &       0     &   0   &  B_2\\
\label{Og}
\end{pmatrix},
\end{equation}
where $\bar{A}$ denotes the conformal quantity.
We will show that the conformal factor does not preserve the algebraic structure of the Pleba{\'n}ski Killing tensors. As a consequence of the deformation, the $K_k$ tensor is no longer a Killing tensor. However, we can define a Conformal Killing Tensor 

\begin{equation*}
    k_\alpha = \frac{1}{6} (2\nabla_\beta \bar{K}^\beta_{\;\;\alpha} + \nabla_\alpha \bar{K}^\beta_{\;\;\beta}),
\end{equation*}

that satisfies the conformal Killing equation 
\begin{equation*}
\bar{K}_{(\mu \nu;\alpha)} = g_{(\alpha \mu} k_{\nu)}.    
\end{equation*}

The same condition is valid for the tensors $\bar{K}_h$ and $\bar{K}_{kh}$ (\ref{KTs}). In a similar procedure to the one in  Sec 2.1, we can compute the Conformal Killing tensor directly
\begin{equation}
 \bar{K}_k^{\mu \nu}= (A_1 + B_1)^{-1}
\begin{pmatrix}
B_1A_5 - A_1B_5 &  B_1A_4 - A_1B_4&   0   &   0\\
B_1A_4 - A_1B_4 &  B_1A_3 - A_1B_3&   0   &   0\\
       0        &         0       & A_2B_1&   0\\
       0        &         0       &   0   &-B_2A_1\\
\end{pmatrix}
\end{equation}
In addition, the conformal tensors $K_h$, $K_{kh}$ can be determined similarly,
\begin{equation}
 \bar{K}_h^{\mu \nu}= (A_1 + B_1)^{-1}
\begin{pmatrix}
A_1A_5 - B_1B_5 &  A_1A_4 - B_1B_4&   0   &   0\\
A_1A_4 - B_1B_4 &  A_1A_3 - B_1B_3&   0   &   0\\
       0        &         0       & A_1A_2&   0\\
       0        &         0       &   0   &-B_1B_2\\
\end{pmatrix}
\end{equation}
\begin{equation}
 \bar{K}_{kh}^{\mu \nu}=  (A_1B_1)^{1/2}(A_1 + B_1)^{-1}
\begin{pmatrix}
A_5 + B_5 &   A_4 + B_4 &   0   &   0\\
A_4 + B_4 &   A_3 + B_3 &   0   &   0\\
    0     &       0     &  A_2  &   0\\
    0     &       0     &   0   &  B_2\\
\end{pmatrix}
\end{equation}
the relation $\Omega^{2}(A_1 - B_1)\bar{\textbf{g}} =  \bar{K}_h-\bar{K}_k$ is easily obtained in matrix form. Let us find the required restrictions on the conformal factor in order to preserve the separability. In this case we have conformal Killing tensors instead Killing tensors, so there is not a fourth conserved quantity.
\subsection{Equations of motion in the  Pleba{\'n}ski metric with conformal factor}

The interaction spacetime-matter is described by means of the trajectories that follow test particles. Solving the HJ equation to determine the motion of a particle is equivalent to solving the geodesic equation:
\begin{equation}
    \Ddot{x}^\alpha + \Gamma^\alpha _{\mu\nu}\Dot{x}^\mu \Dot{x}^\nu = 0,
\end{equation} 
this is a second degree differential equation system in $(\tau,\phi,p,q)$.
\newline
To preserve stationarity and axisymmetry, we assume that $\Omega=\Omega(p,q)$. Then the $\dot{\tau}$ and $\dot{\phi}$ components are found as in the case previously studied from the symmetries on $\tau$ and $\phi$ Eq. (\ref{sol TF}).

The geodesic equation for $\Dot{p}$, after few algebraic manipulations, is
\begin{equation}    
\frac{d}{d\lambda}\left( \frac{(\Omega^2\Sigma)^2\Dot{p}}{P} \right) + \frac{(\Omega^2\Sigma)^2}{P}\frac{d}{d\lambda}\left(\Dot{p}\right)+ \left[\Omega^2\Sigma(\mu^2 + \bar{g}_{ij}\Dot{x}^i\Dot{x}^j) \right]_p =0,
\label{pre p}
\end{equation}
where $i,j=\tau,\phi$, the subscript $p$ denotes the partial derivative with respect to the coordinate $p$; the factor $(\bar{g}_{ij}\Dot{x}^i\Dot{x}^j)_p$ is equal to $(c_\tau p^2\omega + c_\phi)^2/P$ as previously. Eq. (\ref{pre p}) can be simplified as follows
\begin{equation}
 \frac{d}{d\lambda}\left[\frac{(\Omega^2\Sigma)^2\Dot{p}^2}{P} + \frac{(c_\tau\omega  p^2 + c_\phi)^2}{P}\right]+ \mu^2\Dot{p}(\Omega^2\Sigma)_p = 0.
 \label{sol p}
\end{equation}
Analogously for the coordinate $q$, $(\bar{g}_{ij}\Dot{x}^i\Dot{x}^j)_q = - (c_\tau \Xi(q)^2 - c_\phi\omega)^2/e^{2\Phi}Q$, then:
\begin{equation}
   \frac{d}{d\lambda}\left[\frac{(\Omega^2\Sigma)^2\Dot{q}^2}{Q} -\frac{(c_\tau \Xi(q)^2-c_\phi \omega)^2}{e^{2\Phi}Q}\right]+ \mu^2\Dot{q}(\Omega^2\Sigma)_q = 0.
  \label{sol q}
\end{equation}
The obvious condition to impose separability on the HJ equation is that the conformal factor (\ref{Og}) be  $\Omega= $cte or $\Omega^2\Sigma = A_6(p) + B_6(q)$. 
We have preferred a method that explicitly displays how the conformal factor affects the integrability of the system, and thus determines what conditions can be imposed for the system to be integrable. Assuming $\Omega^2\Sigma = A_6(p) + B_6(q)$, we obtain
\begin{equation*}
 \begin{split}
    \dot{p}^2 &
 = -\frac{P A_6(p)}{\Omega^4\Sigma^2}\mu^2 + \frac{PC_1}{\Omega^4\Sigma^2} - \frac{\left(c_{\tau}\omega  p^2  +c_{\phi} \right)^2}{\Omega^4\Sigma^2}, \\
    \dot{q}^2 & = -\frac{Q B_6(q)}{\Omega^4\Sigma^2}\mu^2 + \frac{QC_2}{\Omega^4\Sigma^2} + \frac{\left(c_{\tau}\Xi^2(q)- c_{\phi} \omega \right)^2}{e^{-2\Phi}\Omega^4\Sigma^2},
 \end{split}
\end{equation*}
where $C_1, C_2$ are the integration constants to be determined. The consistency with the case without conformal factor suggests that $A_6 \rightarrow A_1 $ and $B_6 \rightarrow B_1$ if $\Omega \rightarrow 1$. In this limit, the integration constants are related by $C_1=-C_2=C$.
\newline
However, the normalization condition imposes an additional restriction,
\begin{equation*}
\bar{g}^{\mu \nu}\dot{x}_\mu\dot{x}_\nu=-\mu^2/\Omega^2= \frac{A_6 + B_6}{\Omega^2\Sigma}\mu^2 + \frac{C_1 + C_2}{\Omega^2\Sigma},    
\end{equation*}
in such a way that to fulfill the normalization condition $A_6=A_1$, $B_6=B_1$ and $C_1=-C_2$. This implies $\Omega(p,q) = 1$; therefore, to obtain separability and acceptable physical behavior, we return to the non conformal case. 
This implies that for the integration of trajectories of massive test particles, the conformal factor $\Omega(p,q)  $ is inadmissible if we want to preserve the mass invariance. Therefore the introduction of a conformal factor does not preserve the HJ separability. With respect to massless test particles, it is well known that null trajectories are not affected by metric conformal factors. 


\subsection{KG separability for the generalized Pleba{\'n}ski metric with conformal factor}
Let us discuss the implications of the conformal factor in the separability of KG. The introduction of a conformal factor modifies the connections, i.e. the Christoffel symbols transform as
\begin{equation*}
\Gamma^\mu_{\alpha\beta} \rightarrow \Gamma^\mu_{\alpha\beta} + C^\mu_{\alpha\beta} \text{\;where\;}  C^\mu_{\alpha\beta} = \Omega^{-1} (\delta^\mu_\alpha\Omega_\beta + \delta^\mu_\beta\Omega_\alpha - g_{\alpha\beta}g^{\mu \lambda} \Omega_\lambda),
\end{equation*}
in what follows we denote $\nabla_\alpha\Omega = \Omega_{,\alpha} = \Omega_\alpha$. The conformal factor modifies as well the Riemann tensor $\bar{R}$ \cite{Carr},
\begin{equation*}
\bar{R}^\alpha_{\;\;\beta\mu\nu} \rightarrow R^\alpha_{\;\;\beta\mu\nu}  - 2 \Omega^{-1}(\delta^\alpha_{\;[\mu}\delta^{\rho}_{\;\nu]}\delta^\sigma_\beta - g_{\beta[\mu}\delta^\rho_{\;\nu]} g^{\alpha \sigma} ) \Omega_{\rho\sigma} + 2 \Omega^{-2} (2\delta^\alpha_{\;[\mu}\delta^{\rho}_{\;\nu]}\delta^\sigma_\beta -2g_{\beta[\mu}\delta^\rho_{\;\nu]} g^{\alpha \sigma} + g_{\beta[\mu}\delta^\alpha_{\;\nu]} g^{\rho \sigma} )\Omega_\rho\Omega_\sigma,
\end{equation*}
and the Ricci tensor
\begin{equation*}
    \bar{R}_{\mu\nu} \rightarrow R_{\mu\nu} - \Omega^{-1}[2\delta^\alpha_\mu\delta^\beta_\nu + g_{\mu\nu}g^{\alpha\beta}]\Omega_{\alpha\beta} + \Omega^{-2} [4\delta^\alpha_\mu\delta^\beta_\nu -g_{\mu\nu}g^{\alpha\beta}]\Omega_\alpha\Omega_\beta,
\end{equation*}
The conmutator of the $\bar{K},\bar{R}$ to test the anomaly-free condition (\ref{KR}), 
\begin{equation}
    [\bar{K},\bar{R}]^\alpha_{\;\beta} = [K,R]^\alpha_{\;\beta} + \chi R^{\alpha \sigma}K_{\sigma \beta} - \Omega^{-1} \left( 
2[\bar{K},\Omega]^\alpha_{\;\beta} + \Omega^\mu_{\;\mu} \chi g^{\alpha \sigma}K_{\sigma \beta} \right)+ \Omega^{-2} \left( 4 [\bar{K},\Omega\Omega]^\alpha_{\;\beta} -\Omega^\mu\Omega_{\mu}\chi g^{\alpha \sigma}K_{\sigma \beta}\right);
\label{Kbar}
\end{equation}
where we have denoted $\chi = (1-\Omega^4)$; the first term is the commutator in (\ref{KR}) discussed before; since the commutators $[\bar{K},\Omega]^\alpha_{\;\beta} = \bar{K}^{\alpha \sigma}\Omega_{\sigma \beta} - \Omega^{\alpha \sigma} \bar{K}_{\sigma \beta}$ and $[\bar{K},\Omega\Omega]^\alpha_{\;\beta} = \bar{K}^{\alpha \sigma}\Omega_{\sigma}\Omega_{\beta} - \Omega^{\alpha}\Omega^{\sigma} \bar{K}_{\sigma \beta}$, \newline
we see that the introduction of the conformal factor changes the nature of the $K_{\mu \nu}$ tensor, in the sense that it is no longer a Killing tensor, but a conformal Killing Tensor. From the above expressions, it is difficult to guarantee that the fulfillment of (\ref{Kbar}) implies the separability of KG, and additional criteria are needed. A separability criterion was presented in \cite{Papadp}, where it states that the radial and the angular parts separate if and only if:
\begin{equation}
\eta = \ln{\left( \frac{\sqrt{-\text{det}g}}{A_1+B_1} \right)} = \mathcal{F}_1(p) + \mathcal{G}_1(q)
\end{equation}
Where $\mathcal{F}_1(p)$, $\mathcal{G}_1(q)$ are functions that depend only on their arguments. In our case
\begin{equation}
    \eta = \ln{\left( \frac{\sqrt{-\text{det}g}}{A_1+B_1} \right)} = \ln{\left( \frac{\sqrt{\Omega^8e^{-2\Phi}(A_1+B_1)^2}}{A_1+B_1} \right)} =  \ln{\left( \Omega^4e^{-\Phi}\right)} = 4\ln{\Omega} -\Phi(q) 
\end{equation}
this imposes a restriction on the form of the conformal factor, $\Omega = A_6(p)B_6(q)$. In contrast to the  separability of HJ, that demands $\Omega = A_6(p)+B_6(q)$. 
 \section{Sources for the generalized Pleba{\'n}ski metric}\label{sect5}
We conclude the discussion by analyzing the relationship between the generalized functions and the material fields required for the generalized metric to satisfy the Einstein equations. 
In \cite{Lobo} the idea of reconstructing sources to bouncing spacetimes has been explored. In close relations is the work in \cite{Alencar}, where general spherically symmetric black bounces sourced by non-linear electromagnetic fields are studied.

For the purposes of illustration, we consider as sources a scalar field and a nonlinear electromagnetic field, with the action

\begin{equation*}
S= \int \sqrt{-g}\;d^4x\left[\frac{R}{2} + \epsilon g^{\mu \nu}\partial_\mu\Theta\partial_\nu\Theta-V(\Theta)-L(F) \right],
\end{equation*}
where $\epsilon = \pm 1 $ allows us to consider positive or negative kinetic energy for the scalar field; the electromagnetic Lagrangian, $L(F)$, is a function of the electromagnetic invariant $F$ that can correspond to nonlinear corrections of Maxwell's electromagnetic field.   Varying the action, the equations of motion derived are
\begin{equation}
\nabla_\mu[L_F F^{\mu\nu}] = 0,
\end{equation}
\begin{equation}
2\epsilon  \nabla_\mu  \nabla^\mu \Theta = - \frac{dV(\Theta)}{d\Theta},
\end{equation}
\begin{equation}
G_{\mu\nu} = R_{\mu\nu} - \frac{1}{2}g_{\mu\nu}R = T_{\mu\nu}= (T^{\Theta}_{\mu \nu} + T^{EM}_{\mu\nu}),
 \end{equation}
where $T^{\Theta}_{\mu \nu}$ and $T^{EM}_{\mu\nu}$ correspond to the scalar and electromagnetic stress-energy tensors, respectively,   
\begin{equation}
T^{\Theta}_{\mu \nu} = 2\epsilon\partial_\mu\Theta\partial_\nu\Theta-g_{\mu\nu}(\epsilon\partial^\alpha\Theta\partial_\alpha\Theta-V(\Theta));\;\; T^{EM}_{\mu\nu} = g_{\mu\nu}L(F) - L_FF_{\mu \alpha}F_\nu^{\;\alpha},
\label{Ts}
\end{equation}
in case $L(F)=F$ Maxwell electrodynamics is recovered. 
Our goal is to write down the unknown functions in Eq. (\ref{Ts}),  $\Theta$,  $V(\Theta)$, $L$ and $L_F$ in terms of the metric components in Eq. (\ref{PD4}); we shall restrict ourselves to the static case, i.e. as in (\ref{PD4}) with $\omega = 0$.
  
Let us consider the static spherically symmetric metric given by:
\begin{equation}
    ds^2 = -f_1(r)dt^2+ f_2(r)dr^2 + \Xi(r)^2 \left( d\theta^2 + \sin^2\theta d\varphi^2\right),
    \label{spff}
\end{equation}
where $f_1(r)$, $f_2(r)$ and $\Xi(r)$ are arbitrary functions on the radial coordinate $r$. The metric (\ref{spff}) is obtained from (\ref{spherical}) with $p \mapsto \cos{\theta}, \quad q \mapsto r, \quad \tau \mapsto t$ and $\phi \mapsto \varphi$, also making $f_1 = e^{-2 \Phi}/f_2, \quad f_2= \Xi(r)/ Q(r)$ and $P(p) \mapsto (1-p^2)$.

Considering only the electric field, the nonvanishing component of the electromagnetic tensor is $F^{01} = - F^{10}$, while the electromagnetic invariant is
$F= - Q_e^2/(2f_1f_2 L_F^2 \Xi^4)$. 
On the other hand, for a magnetically charged metric, the nonvanishing electromagnetic component is $F^{32}=  Q_m^2/(2f_1f_2 \Xi^4) $ and 
it is convenient to define the auxiliary tensor $P_{\mu\nu}$ as $P_{\mu\nu}=L_FF_{\mu\nu}$, analogously to the constitutive equations in an electromagnetic medium. Then the invariant is $P=P^{\mu\nu}P_{\mu\nu}/4=L_F^2F^{\mu\nu}F_{\mu\nu}/4=L_F^2F= Q_m^2/2r^4$, so that the radial coordinate in terms of $P$, is given by $r^4= Q_m^2/2P$ and the Lagrangian can be written as a function of $L(P)$.

For the static spherically symmetric metric it is convenient to consider $\Theta= \Theta (r)$;  the unknown functions  $\Theta$,  $V(\Theta)$, $L$ and $L_F$ in terms of $f_1(r) $, $ f_2(r)$ and $\Xi (r)$ are given by
\begin{equation}
\begin{split}
\Theta'(r) &= \sqrt{\frac{-2f_1f_2\Xi'' + \Xi'\left(f_1f_2 \right)'}{2\epsilon\Xi f_1f_2}},\\
\frac{dV(r)}{dr} & = - \Theta'(r)\frac{ 2\epsilon}{(f_1f_2)^{1/2}\Xi^2} \partial_r \left[(f_1/f_2)^{1/2}\Xi^2 \partial_r\Theta\right], \\
L_F(r) &= -\frac{4f_1f_2 Q_e^2}{\Xi^2 \left[ 2f_1''f_1f_2\Xi^2 +4f_1^2f_2^2 - 4f_1^2f_2(\Xi'^2 + \Xi'' \Xi) - f_1'(f_1f_2)'\Xi^2  +2f_1(f_1f_2)'\Xi \Xi'\right] }, \\
L(r) & = -V(r) + \frac{2f_1''f_1f_2\Xi - \Xi f_1'(f_1f_2)' +4 f_1f_1'\Xi'f_2}{4\Xi f_1^2f_2^2}.
 \end{split}
 \label{VLLf}
\end{equation}
Here, our task is completed, the fields that support the general metric (\ref{PD4}) in the static case are determined. The components of the energy-momentum tensor $T_{\mu \nu}$
to plug into the Einstein equations are given by
\begin{equation}
 \begin{split}
         T^0_{0} - T^1_{1} = & -2\epsilon \frac{\Theta'^2}{f_2}\\
         T^0_{0} + T^1_{1} = & 2 \left(L + \frac{q^2}{f_1f_2\Xi^4L_F} +  V \right)\\
         T^0_{0} - T^2_{2} = & \frac{q^2}{f_1f_2\Xi^4L_F}
 \end{split}
 \label{Ttt}
\end{equation}
\newline
And in terms of the metric functions in (\ref{spherical}), $\Phi(r)$ and $Q(r)$, which correspond to $f_1(r)=e^{-2\Phi(r)}/f_2(r)$ and $f_2=\Xi^2/Q(r)$; with these substitutions the scalar and electromagnetic fields are defined by 
\begin{equation*}
 \begin{split}
      \Theta'(r) &= \sqrt{\frac{-2e^{-2\Phi(r)}\Xi'' + \Xi'[e^{-2\Phi(r)}]'}{2\epsilon\Xi e^{-2\Phi(r)}}} = \sqrt{\frac{-\Xi'' - \Phi'\Xi'}{\epsilon\Xi }}\\
      \frac{dV(r)}{dr} & = - \Theta'(r)\frac{ 2\epsilon}{e^{-\Phi(r)}\Xi^2} \partial_r \left[e^{-\Phi(r)}\Xi^2 \partial_r\Theta\right], \\
      L_F(r) &=  -\frac{2 Q_e^2e^{2\Phi(r)}}{ \Xi^2(2\Phi'^2Q-3\Phi'Q'-2Q\Phi''-Q''+2) +4Q\Xi (\Xi'\Phi'-\Xi'') -4\Xi \Xi'Q'+ 4\Xi''Q }, \\
      L(r) & = -V(r) + \frac{-2\Xi''\Xi Q-\Xi^2(2Q\Phi''-Q''-2Q\Phi'^2+3\Phi'Q')+2(Q\Phi'-Q')\Xi'\Xi +2Q\Xi'^2}{2\Xi^4},
 \end{split}
\end{equation*}

The generalized functions in metric (\ref{spherical}) contain interesting metrics as particular cases, some of them are listed in
table \ref{tab1}. Moreover the generalized functions allow the introduction of additional parameters, like a nonlinear electromagnetic parameters or a Lorentz symmetry breaking parameter, as we shall see in the following examples.

\begin{table}[h!]
\centering
 \begin{tabular}{|c||c c c |}
 \hline
Spacetime & $Q(r)$ & $\Phi(r)$ & $\Xi(r)$ \\
 \hline \hline
 Static spherical symmetry  &   arbitrary                  &arbitrary&arbitrary      \\
 Schwarzschidl              & $1-\frac{2m}{r}$              &   0   & $r$            \\
 Reissner-Nordstrom         &$1-\frac{2m}{r}+\frac{q^2}{r^2}$&  0   & $r$            \\
 Kiselev                    &   arbitrary                  &    0   & $r$            \\
 Hayward   \cite{Hywrd}     & $1-2mr^2/(r^3+2lm)$        &  0   &$r$             \\
 Breton-Galindo NLE \cite{BrGa}& $1-\frac{2m}{r}+\frac{q^2}{r^2}(1+\zeta r^3)$&0&$r$             \\
 Exponential solution \cite{Yilmaz1958,Yilmaz1973,Misner1999,Robertson1999,BenAmots2007, BenAmots2011}    & $e^{-2m/r}$                        &   0    &$re^{m/r}$      \\
 Morris-Thorne wormhole     & $r^2+l^2$                    &   0    &$\sqrt{r^2+l^2}$\\
 Simpson-Visser black bounce& $r^2+l^2-2m\sqrt{r^2+l^2}$   &   0    &$\sqrt{r^2+l^2}$\\
Kalb-Ramond LSB \cite{Yang}& $r/(l+1)-2m$                 &   0    &$\sqrt{r^2+l^2}$ \\
Bumblebee LSB   \cite{Casana}& $(r-2m)/(1+l)$   & $\ln{(l+1)^{-1/2}}$ &$g\sqrt{r^2+l^2}$\\
Q-Schwarzschild LSB  \cite{QtmSch}& $(r-2m)/(1+l/r)$ & $\ln{(l/r+1)^{-1/2}}$ &$g\sqrt{r^2+l^2}$\\
 \hline \hline
 \end{tabular}
\caption{Various spacetimes included in the generalized Pleba\'nski class discussed herein. }
\label{tab1}
\end{table}
If we examine what happens when the polynomial in $p$ is an arbitrary (well-behaved) function, as in the line element(\ref{spherical})
\begin{equation*}
   ds^2= -f_1(r)\textbf{d}t^2 + f_2(r)\textbf{d}q^2 + \Xi(r)^2\left[P(p)\textbf{d}\phi^2 + \frac{1}{P(p)}\textbf{d}p^2 \right].
\end{equation*}
The previous expressions remain the same, the only difference being the Lagrangian derivative,
\begin{equation*}
     L_F (r)= -\frac{4f_1f_2 Q_e^2}{\Xi^2( 2f_1'' f_1f_2\Xi^2 -2P''f_1^2f_2^2 - 4f_1^2f_2(\Xi'^2 + \Xi'' \Xi) - f_1'[f_1f_2]'\Xi^2  +2f_1[f_1f_2]'\Xi \Xi') }.
\end{equation*}
The general functions have been consistently integrated with the sources for the static case ($\omega=0$).

\subsection{Dominant Energy Condition for the generalized Pleba{\'n}ski metric}
Usually, a source can be found for a given geometry, however,  to check if the proposed matter is physically reasonable, we must at least show that it obeys sensible energy conditions. To this end we analyze the restrictions that the dominant energy condition impose to the generalized or distorted Pleba{\'n}ski metric functions.

The general functions $A_i,B_i$ in terms of the spherical metric (\ref{spff})  are
\begin{equation*}
 \begin{split}
     A_1 & = 0, \;\;\; A_2 = 1, \;\;\; A_3 = 1/\sin^2{\theta}, \;\;\; A_4 =0, \;\;\; A_5 = 0\\
     B_1 & = \Xi(r)^2, \;\; B_2 =\Xi(r)^2/f_2(r), \;\;\; B_3 = 0, \;\;\; B_4 = 0, \;\;\; B_5 =- \Xi(r)^2/f_1(r).
 \end{split}
\end{equation*}
and the components of the 4-velocity of a test particle
derived from Eqs. (\ref{sol TF}), (\ref{upq}), are
\begin{equation*}    
u^{\mu} = \left( -\frac{c_\tau}{f_1(r)},\;\; \frac{c_\varphi}{\Xi^2\sin^2\theta},\;\;\sqrt{\frac{\Xi^2f_1\mu^2+c_\tau^2\Xi^2+Cf_1}{\Xi^2f_1f_2}},\;\; \frac{\sqrt{C\sin^2\theta-c_\varphi^2}}{\Xi^2 \sin\theta}  \right),
\end{equation*}
where $c_{\tau}$, $c_{\varphi}$ and $C$ are constant;
contracting with $T_{\mu\nu}$ we obtain
\begin{equation*}
T_{\mu\nu}u^\mu u^\nu = \frac{-2\epsilon\Phi'^2\Xi^4L_F \left( Cf_1-\Xi^2(c_\tau^2+f_1\mu^2/2) \right) + \Xi^6f_1f_2\mu^2L_F (L+V) + Q_e^2(\Xi^2\mu^2-C)}{\Xi^6f_1f_2L_F};
\end{equation*}
and in terms of the metric functions in (\ref{spherical}),
\begin{equation*}
\begin{split}
T_{\mu\nu}u^\mu u^\nu = &\frac{4f_1f_2\Xi\Xi''(Cf_1-2c_\tau^2\Xi^2)+ 2Cf_1f_2\Xi^2f_1'' +4f_1^2f_2(\Xi^2\mu^2-C)\Xi''}{4f_1^2f_2^2\Xi^4}  \\
&\;\;\;- \frac{2\Xi\Xi'
\left( f_2f_1'(-2\Xi^2(c_\tau^2+f_1\mu^2)+Cf_1) + f_1f_2'\Xi'(Cf_1-2c_\tau^2\Xi^2) \right)}{4f_1^2f_2^2\Xi^4} \\
& + \;\;\; \frac{4f_1^2f_2^2(C-\mu^2\Xi^2) - Cf_1'\Xi^2(f_1f_2)'}{4f_1^2f_2^2\Xi^4};
\end{split}
\end{equation*}
for the particular case $f_1=f_2^{-1}$ we have
\begin{equation*}
    T_{\mu\nu}u^\mu u^\nu = \frac{2f_2\Xi\Xi''(C-2c_t^2f_2\Xi^2) - Cf_2\Xi^2 f_2'' + 2f_2^2(C-\Xi^2\mu^2)(\Xi'^2-1) -2\mu^2f_2\Xi^3f_2'\Xi'+2C\Xi^2f_2'^2}{2\Xi^4f_2^3}.
\end{equation*}

In the massless case, $\mu=0$
\begin{equation*}
    T_{\mu\nu}u^\mu u^\nu = \frac{-2\epsilon\Phi'^2\Xi^4L_F\left( Cf_1-\Xi^2c_\tau^2 \right) -C Q_e^2}{\Xi^6f_1f_2L_F};
\end{equation*}
and in the particular case $f_1=f_2^{-1}$
\begin{equation*}
    T_{\mu\nu}u^\mu u^\nu = \frac{2f_2\Xi\Xi''(C-2c_t^2f_2\Xi^2) + 2Cf_2^2(\Xi'^2-1) + C\Xi^2(2f_2'^2 - f_2 f_2'')}{2\Xi^4f_2^3}.
\end{equation*}
The dominant energy condition (DEC) demands that
$T_{\mu\nu}u^\mu u^\nu \ge 0$, i.e. that mass-energy can never be observed to be flowing faster than light. In the next subsection, we shall check this condition in the considered examples.


\subsection{Examples}
Now we present some examples to illustrate the wide range of situations comprised in the generalized PD metric and how these metrics can be sourced, i.e. the resultant geometry can be sourced in a non-unique form. First, we consider metrics such that $\Xi(r)=r$ and $f_1(r)=f_2^{-1}(r)$. The condition
$\Xi(r)=r$ leads to the absence of a scalar field (consequently $V(\Theta)=0$),  then the matter in these cases is only electromagnetic.
\newline

\subsubsection{NLE generalization of Reissner-Nordstrom}

We consider the metric (\ref{spff}) with $f_1(r)=f_2^{-1}(r) = f(r)$, and $f(r)$ corresponding to the  Reissner--Nordstrom metric plus an arbitrary function $Q(r)$,
\begin{equation*}
f(r) = 1 -\frac{2m}{r}+\frac{Q_e^2}{r^2} + Q(r),
\end{equation*}
where $m$ is the mass and $Q_e$ is the electric charge of the BH.
It is known that the Reissner-Nordstrom metric is the only static, spherically symmetric and asymptotically flat solution of the Einstein-Maxwell equations, see \cite{Heusler}. 
If we impose that $Q(r) \approx r^{-2}$ at infinity then asymptotic flatness is preserved. In this case, the presence of $Q(r)$ modifies the right-hand side of the Einstein equations, implying that a source, in addition to Maxwell matter, is required; it turns out that the source can be a nonlinear electromagnetic field derived from the Lagrangian,
\begin{equation*}
\begin{split}
L&=  \frac{Q_e^2}{r^4}+ \frac{Q' }{r} + \frac{Q''}{2}, \\
L_F &= -\frac{2Q_e^2}{Q''r^4 -2Qr^2+4Q_e^2}.
\end{split}
\end{equation*}
The energy flux measured by an observer with four-velocity
$u^{\mu}$,
\begin{equation}
T_{\mu\nu}u^\mu u^\nu  = 
\frac{2C  Q_e^{2}}{ r^{6}} +
\frac{C \left(Q''r^{2} -2Q\right)}{2 r^{4}};    
\end{equation}
such that the dominant energy condition is satisfied if $C >0$ and
$(Q''r^{2} -2Q) > 0$.

A particular case is $Q(r)= Q_e^2 \zeta r$, where $\zeta$ is a NLE parameter, that has a metric function given by  
\begin{equation*}
f(r) = 1 -\frac{2m}{r} + \frac{Q_e^2}{r^2}(1+\zeta r^3).
\end{equation*}
This metric becomes a solution to the Einstein equations sourced by a nonlinear electromagnetic (NLE) field. Let us consider a magnetic charge, $Q_m$, instead of an electric one, in order to give a closed expression of the Lagrangian in terms of the electromagnetic invariant $P= Q_m^2/r^4$,
\begin{equation*}
\begin{split}
L(r) & = \frac{Q_m^2}{r^4} \left\{ 1 +  \zeta r^3 \right\},\\
L(P) & = 2P \left[ 1 +  \zeta \left(\frac{Q_m^2}{2P} \right)^{3/4} \right],\\
L_F(r) &= \frac{1}{\zeta r^3-2}  \\
F(r) & =- \frac{  Q_m^2}{2r^4} \left[ \zeta r^3 - 2\right]^2.
\end{split}
\end{equation*}
The linear limit of the solution is the magnetized Reissner-Nordstrom and is recovered if $\zeta=0$, and $P=F= Q_m^2/r^4$.
For the electromagnetic stress-energy tensor projected on the four-velocity of an observer,
\begin{equation}
T_{\mu\nu}u^\mu u^\nu  = \frac{ C Q_e^{2}}{r^6} 
\left\{ 2 - \zeta r^3 \right\},    
\end{equation}
and  if $C > 0$ and the nonlinear parameter $\zeta < 0$, then reasonable energy conditions (DEC) are met.
In \cite{BrGa} an analogous metric was derived;  in that work, the authors used a method based on adjusting the electromagnetic potential $A_{\mu}$ to obtain a solution. In contrast, we approach the problem differently, without modifying the potential; in fact, we do not even require an explicit expression for it.

\subsubsection{The Hayward metric}
 
The Hayward metric \cite{Hywrd} is a model of a regular black hole, i.e., a black hole solution to the Einstein field equations that avoids curvature singularities at the center. It was proposed by  Hayward in 2005 as a modification of the Schwarzschild solution that remains finite and well defined at $r=0$, addressing one of the key physical issues in classical general relativity, that is,  the singularity problem.
The metric function in the line element (\ref{spff}),  with 
$f_1(r)=f_2^{-1}(r) = f(r)$, is given by 
\begin{equation*}
f(r) = 1 -\frac{2mr^2}{r^3+2l^2m},
 \end{equation*}
where $l$ is a convenient encoding of the central energy density $3/8\pi l^2$,  assumed to be positive,  often associated with quantum gravity effects or a fundamental length scale \cite{Rvll}.
As a source for this geometry we obtain the NLE field characterized by
\begin{equation*}
\begin{split}
L_F(r) &=  -\frac{q^2(2l^2m+r^3)^3}{36r^7m^2l^2 } \\
L(r) & =   -\frac{24m^2l^2(l^2m-r^3)}{(2l^2m+r^3)^3}\\
F(r) &= -\frac{648r^{10}m^4l^4}{q^2(2l^2m+r^3)^6}.
\end{split}
\end{equation*}
With respect to the energy condition DEC, we obtain the following,
\begin{equation}
T_{\mu\nu}u^\mu u^\nu = \frac{36 r C \,m^{2} l^{2}}{\left(2 l^{2} m+r^{3}\right)^{3}},    
\end{equation}
then if the Carter constant is positive, $C > 0$, the dominant energy condition is satisfied.

Some other examples are studied in \cite{Saez}. To include a scalar field contribution, in the following examples we modify the dependence of the function $\Xi(r)$ on $\Xi^2=r^2+a^2$, which constitutes a Black-Bounce-like modification. \newline
\subsubsection{Lorentz symmetry breaking metrics}
The following examples are related to spacetimes characterized by Lorentz symmetry breaking (LSB) measured by the parameter $l$; although presented in the literature as unconnected cases, these metrics can be sourced from very similar matter, as we show in what follows.

These metrics are characterized by $\Xi(r)^2 = r^2 + l^2$ that is a deformation of the radial coordinate sometimes called the {\it bounce}, which is a way of removing the singularity at the origin, $r=0$. These metrics can be sourced by scalar and NLE fields
and it is the scalar field that precisely supports the bounce through a phantom scalar field. However, the potential associated to the scalar field turns out to be real.

\subsubsection{Kalb-Ramond (KR) field}
A static and spherically symmetric black hole in gravity with a Kalb-Ramond (KR) field is presented in \cite{Yang}, in the framework of a nonminimally coupled Kalb-Ramond field that breaks the Lorentz symmetry of gravity when acquires a nonzero vacuum expectation value.  The assumed geometry is given by
\begin{equation}
     ds^2= -\left(\frac{1}{1-l}-\frac{2m}{r}\right)dt^2 + \left(\frac{1}{1-l}-\frac{2m}{r}\right)^{-1}dr^2+ \Xi(r)^2(d\theta^2 + \sin^2\theta d\varphi^2),
\label{LSB1}     
\end{equation}
where $m$ is the BH mass and $l$ is the parameter that characterizes the spontaneous breaking of Lorentz symmetry.   In terms of the line element (\ref{spff}), the metric functions are
 \begin{equation*}
      f_1(r) = f_2(r)^{-1} = \left(\frac{1}{1-l}-\frac{2m}{r}\right),\;\;\; \Xi(r)^2=r^2+l^2;
 \end{equation*}
the metric (\ref{LSB1}) can be sourced by the scalar field, 
\begin{equation}
\begin{split}
     \Theta(r) & = \frac{1}{\sqrt{-\epsilon}} \tan^{-1}\left({\frac{r}{l}}\right)  \\
     V(r) &= -\frac{2m}{l{^3}} \left[ 3\tan^{-1}{\left(\frac{r}{l}\right)} + \frac{3lr^2 +2l^3}{(r^2+l^2)r}  \right]= -\frac{2m}{l{^3}} \left[ 3 {\sqrt{-\epsilon}} \Theta(r)  + \Theta(r)^{\prime}  (3 l \tan \left( {\sqrt{-\epsilon}} \Theta(r) \right) +2l^2) \right]
\end{split}
\label{sfLSB1}
\end{equation}
and the nonlinear electromagnetic field given by the Lagrangian
\begin{equation*}
\begin{split}     
      L(r) & = \frac{6m}{l{^3}}  \tan^{-1}{\left(\frac{r}{l}\right)} + \frac{2m}{(r^2+l^2)} \left( \frac{2l^2+3r^2}{l^2r} - \frac{l^2}{r^3} \right)  \\
      L_F(r) &=  \frac{Q_e^2r^3(l-1)}{(r^2+l^2)\left(2ml^2(l-1) - lr^3\right) } \\
      F(r) & = -\frac{\left(2ml^2(l-1) - lr^3\right)^2}{2Q_e^2r^6(l-1)^2}.
      \end{split}
\end{equation*}
Apparently the Lagrangian $L$ does not depend on the electric charge, $Q_e$, however, it is contained in the coordinate $r$ that from the expression of $L_F$ or $F$ depends on the BH parameters, $m$, $l$ and $Q_e$. 
In this case, the dominant energy condition amounts to 
\begin{equation*}
\begin{split}  
      T_{\mu\nu}u^\mu u^\nu & = \frac{\left(l-1\right) \left(c_\tau^{2}+2 C\right) (r^{4} + l r^3)+ C r^{3} -2Cm(l-1)(r^2+4lr+2l^2)}{2 \left(r+l\right)^{3} r^{3} \left(-1+l\right)}\ge 0,
\end{split}
\end{equation*}
that is fulfilled if $ C$ is positive.
Although the scalar field is imaginary, due to the factor
$\sqrt{- \epsilon}$, it is still possible to fulfill DEC.

\subsubsection{Einstein-Hilbert-Bumblebee BH}

In \cite{Casana} a static spherically symmetric exact vacuum solution from the gravity sector contained in the minimal standard-model extension, is derived, assuming a Riemann spacetime coupled to the bumblebee field which is responsible for the spontaneous Lorentz symmetry breaking. The metric turns out to be
a Schwarzschild-like BH, 
\begin{equation}
ds^2= -\left(1-\frac{2m}{r}\right)dt^2 + (1+l)\left(1-\frac{2m}{r}\right)^{-1}dr^2+g^2\Xi(r)^2(d\theta^2 + \sin^2\theta d\varphi^2),
\label{LSB2}
\end{equation}
In terms of  (\ref{spff}) the metric functions are,
\begin{equation*}
f_1(r) = \left(1-\frac{2m}{r}\right),\;\;\; 
f_2(r) = (1+l)\left(1-\frac{2m}{r}\right)^{-1},\;\;\; \Xi(r)^2=r^2+l^2;
\end{equation*}
The metric (\ref{LSB2}) represents a purely radial Lorentz symmetry breaking solution outside a spherical body characterizing a modified BH.
This spacetime can be sourced by the scalar field and potential, respectively, 
\begin{equation}
\begin{split}
\Theta(r) & = \frac{1}{\sqrt{-\epsilon}} \tan^{-1}\left({\frac{r}{l}}\right) \\
V(r) &= -\frac{2m}{l{^3}(l+1)} \left[ 3\tan^{-1}{\left(\frac{r}{l}\right)} + \frac{3lr^2 +2l^3}{(r^2+l^2)r}  \right]
\end{split}
\label{sfLSB2}
\end{equation}
along with the nonlinear electromagnetic field, 
\begin{equation*}
\begin{split}
L_F(r) &= \frac{q^{2} r^{3}}{ g^{2} \left(\left(g^{2}-l-1\right) r^{3}+2l^{2} m \,g^{2}\right) \left(l^{2}+r^{2}\right)}\\
L(r) & =\frac{2m \left(3 \arctan\! \left(\frac{r}{l}\right) r^{3}-l^{3}+3 r^{2} l\right)}{l^{3} \left(1+l\right) r^{3}}\\
F(r) & = \frac{ {\left(\left(g^{2}-l-1\right) r^{3}+2l^{2} m \,g^{2}\right)}^{2}}{2q^{2} r^{6} \left(1+l\right)}
\end{split}
\end{equation*}

 While the dominant energy condition amounts to
\begin{equation*}
\begin{split}      
T_{\mu\nu}u^\mu u^\nu & = \frac{\left(\left(2 l+2\right) C+g^{4} \mathit{c_\tau}^{2}\right) r^{4}+\left(\left(2 l^{2}-g^{2}+2 l\right) C+l \,g^{4} \mathit{c_\tau}^{2}\right) r^{3}-2 C m \,g^{2} (r^{2} + 4 l  r + 2 l^{2})}{2 r^{3} g^{4} \left(1+l \right) \left(r+l\right)^{3}} \ge 0,
\end{split}
\end{equation*}
and then DEC may be fulfilled with convenient motion constants $C$, since $c_\tau^2 > 0$.
Note that curiously the scalar fields in these two last cases, 5.2.4 and 5.2.5 are the same, cf. Eqs. (\ref{sfLSB1}) and (\ref{sfLSB2}).

In the last examples, taking the limit $l \rightarrow 0$ leads to $\Xi(r) \rightarrow r$. Although the solutions approach the Schwarzschild metric, implying that the sources vanish, this transition disrupts the consistency of the proposed model in associating the sources with the general functions. We believe that this issue reveals a limitation in our proposed Lagrangian, as it does not account for non-minimal coupling between the scalar and electromagnetic field, whereas these metrics are derived from an interaction-based model \cite{Bumblebee}.\newline
All of them obeys $L_{,r}-L_{,F}F_{,r}=0$. But the implicit definitions of $L(r)$ and $L_F(R)$ in (\ref{VLLf}) do not fulfill that requirement. The next subsection illustrates this point.

\subsection{Quantum Schwarzschild }
An effective theory to describe the quantization of spherically symmetric vacuum motivated by loop quantum gravity has been presented in \cite{QtmSch}. The quantum Schwarzschild black hole is a static region described by the metric
 \begin{equation*}
     ds^2= -\left(1-\frac{2m}{r}\right)dt^2 + \left(1-\frac{l}{r} \right)^{-1}\left(1-\frac{2m}{r}\right)^{-1}dr^2+\Xi(r)^2(d\theta^2 + \sin^2\theta d\varphi^2).
 \end{equation*}
Quantum-gravity effects introduce a length scale $l > 0$, that defines a minimum of the area of the orbits of the spherical symmetry, and removes the classical singularity. 
In terms of the metric functions in (\ref{spff}),
\begin{equation*}
      f_1(r) = \left(1-\frac{2m}{r}\right),\;\;\; f_2(r) = (1-\frac{l}{r})^{-1}\left(1-\frac{2m}{r}\right)^{-1},\;\;\; \Xi(r)^2=r^2+l^2
\end{equation*}
with $r \in (2m, \infty)$, this region is asymptotically flat, and will describe one exterior domain. Following the previous reasoning the fields that may support this solution are  
\begin{equation*}
\begin{split}
\frac{d}{d r}\Phi \! \left(r\right)&=\frac{1}{\sqrt{\epsilon}}{\left(\frac{l \left(l^{2}-2 l r-r^{2}\right)}{2 \left(l^{2}+r^{2}\right)^{2} \left(r-l\right)}\right)}^{\frac{1}{2}}\\
V'\left(r\right)&=\frac{l \left(r^{3}+2mr^{2} +l^{2} r+8lmr-6l^{2} m\right)}{2 \left(l^{2}+r^{2}\right)^{2} r^{3}}\\
L_F(r) =&\frac{2 \left(l-r\right) r^{3} q^{2}}{\left(5 l^{2} m-4 r l m+3 r^{2} m+r^{3}\right) l \left(l^{2}+r^{2}\right)} \\
L(r) & = \frac{m l \left(5 l^{2}-4 r l+r^{2}\right)}{2 r^{4} \left(l^{2}+r^{2}\right)}-V(r)\\
F(r) & =\frac{l^{2} \left(5 l^{2} m-4 r l m+3 r^{2} m+r^{3}\right)^{2}}{8 \left(l-r\right) r^{7} q^{2}}\\
\end{split}
\end{equation*}
Here, we leave implicit the expressions for the potential and the field. These fields are consistent with the Einstein equations, but in this case the condition $L_{,r} - L_{,F} F_{,r} = 0$ is not true; to satisfy the electromagnetic condition an ad hoc term is required, $L_{,r} - L_{,F} F_{,r} = \hat{V}(r)$. 
If $\hat{V}(r) = 0$, then the electromagnetic Lagrangian is consistent with $L_{,r} - L_{,F} F_{,r} = 0$.
If we substitute the expressions obtained for $L$, $L_F$ and $V$ (see Eq. (\ref{VLLf})), then $\hat{V}(r)$ is given by:
 \begin{multline*}
8H^{3} \Sigma^{2}\hat{V}(r)=-2 H \left(\Sigma'H'-\Sigma\right) \left(2\Sigma'f_1-\Sigma f_1'\right)H''+2 H \Sigma  \left(- \Sigma   H'\right) f_1''\\+\left(-4 \Sigma'' Hf_1+\left(6 \Sigma' f_1-3 \Sigma f_1'\right) H'-2 H f_1' \Sigma' \right) \Sigma H'\\
+4 \left(2\Sigma'' Hf_1 \Sigma +\left(-2f_1 \Sigma \Sigma' +\Sigma ^{2} f_1'\right) H' +2H f_1 \Sigma'^{2}-H^{2}\right) H'
\end{multline*}
Where, $H(r)=f_1f_2$. This factor vanishes if it is constant; this is the reason it did not appear in the previous cases. Suppose that $2\Sigma'f_1-\Sigma f_1'=0$, then:
 \begin{equation*}
\hat{V}(r)=-H'\frac{\left(2H-2\Sigma f_1 \Sigma'' -4f_1 \Sigma'^{2}+\Sigma^{2} f_1''+\Sigma f_1' \Sigma'\right)}{4 H^2 \Sigma^{2}}
 \end{equation*}
Some attempts to find consistent and no trivial functions, $f_1$, $\Sigma$, $H(r)$ may start by fixing $f_1$ or $\Sigma$, solving $2\Sigma'f_1-\Sigma f_1'=0$ and substituting into $\hat{V}$. \newline
A general solution with a fixed $\Sigma(r)$ is $f_1(r) = C\Sigma^2$ and it implies that

\begin{equation*}
H(r) = cte.
\end{equation*}
This condition is satisfied for instance if $f_1 = {\rm const} f_2$.

These examples illustrate the fact that for a given geometry, if it is not vacuum, there are several sources that could generate that given curvature; i.e. other than vacuum solutions of the Einstein equations, the question of unicity of the solutions does not apply, and the sources can come from very different matter field settings.

\section{Conclusions}\label{sect6}
In this work, we analize the stationary axisymmetric spacetime that results from generalizing the metric functions in the Pleb\'anski spacetime (\ref{PD4}). We found that the separability of the HJ and KG equations is preserved, and we present the integration of test particle trajectories in the deformed Pleb\'anski spacetime, considering both, electrically charged and uncharged particles, using the four motion constants.
We also prove that there is a Killing horizon and then define the surface gravity of the generalized metric.

Moreover, we showed that an additional generalization of the Pleb\'anski  metric including a conformal factor breaks some symmetries such that neither the HJ nor the KG separabilities are preserved.

Finally, we present a proposal for the matter that could generate some static generalized Pleb\'anski metrics. It turns out that the static generalized Pleb\'anski metric can be sourced by a scalar field and nonlinear electromagnetic matter. These fields are expressed in terms of the general metric functions that define the metric. Also, the proposed matter satisfy the dominant energy condition, making them viable solutions of the Einstein equations.

A possible extension of this work involves considering an additional term in the action including interaction between the scalar and electromagnetic field, as well as look for sources for the stationary metrics.
\textbf{}{Acknowledgments}
The work of ASA has been sponsored by Conahcyt-Mexico through the Ph. D.  scholarship No. 839787; NB acknowledges financial support from Conahcyt-Mexico through the project CBF2023-2024-811.
{\bf Data Availability Statement}. This manuscript has no associated data or the data will not be deposited. [Authors’ comment: Data sharing is not applicable since no data has been generated.]

\begin{appendices}
\section{Christoffel Symbols}\label{secA}

For a general stationary axisymmetric spacetime $g_{\mu\nu}$ we found the Christoffel simbols $\Gamma^\mu_{\alpha\beta} = (1/2) g^{\mu \lambda}(g_{\alpha\lambda,\beta} + g_{\lambda\beta,\alpha}-g_{\alpha\beta,\lambda})$

\begin{equation*}
\left[\Gamma_{\mu \nu} ^\tau\right] = \frac{1}{2} \left(
\begin{matrix}
0 & 0 & g^{\tau \tau}g_{\tau \tau,p} + g^{\tau \phi}g_{\phi \tau,p} & g^{\tau \tau}g_{\tau \tau,q} + g^{\tau \phi}g_{\phi \tau,q} \\
0 & 0 & g^{\tau \tau}g_{\phi \tau,p} + g^{\tau \phi}g_{\phi \phi,p} & g^{\tau \tau}g_{\phi \tau,q} + g^{\tau \phi}g_{\phi \phi,q}\\
g^{\tau \tau}g_{\tau \tau,p} + g^{\tau \phi}g_{\phi \tau,p} & g^{\tau \tau}g_{\phi \tau,p} + g^{\tau \phi}g_{\phi \phi,p} & 0 & 0 \\
g^{\tau \tau}g_{\tau \tau,q} + g^{\tau \phi}g_{\phi \tau,q} & g^{\tau \tau}g_{\phi \tau,q} + g^{\tau \phi}g_{\phi \phi,q} & 0 & 0
\end{matrix}
\right)
\end{equation*}

\begin{equation*}
\left[\Gamma_{\mu \nu} ^\phi\right] = \frac{1}{2} \left(
\begin{matrix}
0 & 0 & g^{\phi \tau}g_{\tau \tau,p} + g^{\phi \phi}g_{\phi \tau,p} & g^{\phi \tau}g_{\tau \tau,q} + g^{\phi \phi}g_{\phi \tau,q} \\
0 & 0 & g^{\phi \tau}g_{\phi \tau,p} + g^{\phi \phi}g_{\phi \phi,p} & g^{\phi \tau}g_{\phi \tau,q} + g^{\phi \phi}g_{\phi \phi,q}\\
g^{\phi \tau}g_{\tau \tau,p} + g^{\phi \phi}g_{\phi \tau,p} & g^{\phi \tau}g_{\phi \tau,p} + g^{\phi \phi}g_{\phi \phi,p} & 0 & 0 \\
g^{\phi \tau}g_{\tau \tau,q} + g^{\phi \phi}g_{\phi \tau,q} & g^{\phi \tau}g_{\phi \tau,q} + g^{\phi \phi}g_{\phi \phi,q} & 0 & 0
\end{matrix}
\right)
\end{equation*}
\begin{equation*}
\left[\Gamma_{\mu \nu} ^p \right] = \frac{g^{pp}}{2} \left(
\begin{matrix}
-g_{\tau \tau,p}& -g_{\tau \phi,p} & 0        & 0 \\
-g_{\phi \tau,p}& -g_{\phi \phi,p} & 0        & 0\\
0               & 0                & g_{pp,p} & g_{pp,q} \\
0               & 0                & g_{pp,q} & -g_{qq,p}
\end{matrix}
\right) \; \; \; \;
\left[\Gamma_{\mu \nu} ^q \right] = \frac{g^{qq}}{2} \left(
\begin{matrix}
-g_{\tau \tau,q}& -g_{\tau \phi,q} & 0        & 0 \\
-g_{\phi \tau,q}& -g_{\phi \phi,q} & 0        & 0\\
0               & 0                & -g_{pp,q} & g_{qq,p} \\
0               & 0                & g_{qq,p} & g_{qq,q}
\end{matrix}
\right)
\end{equation*}
\section{Petrov Clasification}\label{secB}
\begin{table}[h!]
\centering
 \begin{tabular}{|c||c  |}
 \hline
Type & Conditions \\
 \hline \hline
 0   & $\Psi_0=\Psi_1=\Psi_2=\Psi_3=\Psi_4=0$ \\
 I   & $D\not=0$ \\
 II  & $D=0,\;I\not=0,\;J\not=0,\;K\not=0,\;N\not=0$ \\
 III & $D=0,\;I=J=0,\;K\not=0,\;L\not=0$   \\
 N   & $D=0,\;I=J=K=L=0$\\
 D   & $D=0,\;I\not=0,\;J\not=0,\;K=N=0$\\
 \hline \hline
 \end{tabular}
     \caption{Petrov classification}
    \label{tab2}
\end{table}
In this Appendix we show how  the simplest generalization, $P(p)$ and $Q(q)$ as arbitrary functions leads to an algebraically general metric. 

The Lorentz invariants are defined as
\begin{equation*}
 \begin{split}
     I&= \Psi_0\Psi_4-4\Psi_1\Psi_3+3\Psi_2^2 \\
     J&= -\Psi_2^3+\Psi_0\Psi_2\Psi_4+2\Psi_1\Psi_2\Psi_3-\Psi_4\Psi_1^2-\Psi_0\Psi_3^2\\
     D&= I^3-27J^2 \\
     K&= \Psi_4^2\Psi_1-3\Psi_4\Psi_3\Psi_2+2\Psi_3^3 \\
     L&= \Psi_2\Psi_4-\Psi_3^2\\
     N&= 12L^2-\Psi_4^2I
 \end{split}
\end{equation*}
The invariants are given by
\begin{equation*}
 \begin{split}
\Psi_{0}&=-\frac{3 Q p^{4} \omega^{2} }{2 \Sigma^{4}} \psi(p,q)\;,\;\;\;\;\;\;\;\;\;\;\;\;\;\;\;\;\;\;\;\;\;\;\;\;\;\;\;\;\;\;\;\;\;\Psi_{1}=-\frac{3 \sqrt{2PQ}\, q^{2} p^{2} \omega    }{4 \Sigma^{\frac{7}{2}} \sqrt{P q^4- Q \omega^{2} p^4}} \psi(p,q)\\
\Psi_{2}&=\frac{Q \omega^{2} p^{4}+2 P q^{4}}{4 \Sigma^{3} \left(-P q^{4}+Q \omega^{2} p^{4}\right)} \psi(p,q) = -\frac{\Sigma}{6(-Pq^4+Q\omega^2p^4)}\Psi_0+\frac{Pq^4}{2\Sigma^3(-Pq^4+Q\omega^2p^4)}\psi(p,q)\\
\Psi_{3}&=-\frac{3 \sqrt{2PQ}\, q^{2} p^{2} \omega }{8\Sigma^{\frac{5}{2}} \left(-Pq^{4}+Q \omega^{2} p^{4}\right)^{3/2}}\psi(p,q) = -\frac{\Sigma}{2i(Pq^4-Q\omega^2p^4)}\Psi_1\\
\Psi_{4}&=-\frac{Q p^{4} \omega^{2} }{16\left(-P q^{4}+ Q \omega^{2} p^{4}\right)^{2} \Sigma^{2}}\psi(p,q) =\frac{\Sigma^2}{24(-Pq^4+Q\omega^2p^4)^2}\Psi_0\\
 \end{split}
\end{equation*}
\begin{equation*}
    \begin{split}
\psi(p,q) & = -\Sigma^{2} (P''+ Q'')/6-\omega  \Sigma\left(\mathrm{I} q-\omega  p\right) P'+\left(\mathrm{I} p \omega +q\right)\Sigma Q'+2 \left(q -I\omega  p\right)^2 \left(P \omega^{2}-Q\right)
    \end{split}
\end{equation*}
For the PD metric when the condition
\begin{equation*}
P'' + Q'' = -4\lambda \Sigma \longrightarrow P''=-4\lambda p^2\omega^4,\;\; Q''=-4\lambda q^2
\end{equation*}
is imposed, that leads to the type D.
Also note that if $\omega=0$ then only $\Psi_2\not=0$, that falls again in the type D.

It is expected the same situation for a more general case, for instance when $q$ is a general function. From the form of the invariants, stationary metrics belong to the type I, except for th ones in type D.  The strongest restrictions that makes the Pleba{\'n}ski metric of type D is $P'' + Q'' = -4\lambda \Sigma$.

\end{appendices}
\bibliographystyle{unsrt}
 
\end{document}